\documentclass[preprint,showpacs,preprintnumbers,amsmath,amssymb]{revtex4}
\usepackage{graphicx}
\usepackage{dcolumn}
\usepackage{bm}
\usepackage{nicefrac}
\usepackage{units}
\usepackage{amsmath}
\usepackage{pgflibraryarrows}
\usepackage{pgflibrarysnakes}
\usepackage{graphicx}
\usepackage{subfigure}
\usepackage{color}
\usepackage{hyperref}

\begin{document}

\title{Accelerating hydrodynamic description of pseudorapidity density and
the initial energy density in p+p, Cu+Cu, Au+Au, and Pb+Pb collisions at RHIC and LHC}

\author{Jiang Ze-Fang$^{~1,2}$}
\email{jiangzf@mails.ccnu.edu.cn}
\author{Yang Chun-Bin$^{~1,2}$}
\email{cbyang@mail.ccnu.edu.cn}
\author{M\'at\'e Csan\'ad$^{~3}$}
\email{csanad@elte.hu}
\author{Tam\'as Cs\"org\H{o}$^{~4,5}$}
\email{tamas.ferenc.csorgo@cern.ch}

\affiliation{$^1$ Key Laboratory of Quark and Lepton Physics, Ministry of Education, Wuhan, 430079, China}
\affiliation{$^2$ Institute of Particle Physics, Central China Normal University, Wuhan 430079, China}
\affiliation{$^3$ ELTE, E{\"o}tv{\"o}s Lor{\'a}nd University, H-1117 Budapest, P{\'a}zm{\'a}ny P. s. 1/A, Hungary}
\affiliation{$^4$ Wigner Research Centre for Physics, Hungarian Academy of Sciences, H-1525 Budapest 114, P.O.Box 49, Hungary}
\affiliation{$^5$ Eszterh{\'a}zy K{\'a}roly University, K{\'a}roly R{\'o}bert Campus, H-3200, Gy{\"o}ngy{\"o}s, M{\'a}trai \'ut 36, Hungary}

\begin{abstract}
A known class of analytic, exact, accelerating solutions of prefect relativistic hydrodynamics with longitudinal acceleration is utilized to describe results on the pseudorapidity distributions for different collision systems. These results include $dN/d\eta$ measured in p+p, Cu+Cu, Au+Au, and Pb+Pb collisions at RHIC and LHC, in a broad centrality range. Going beyond the traditional Bjorken model, from the accelerating hydrodynamic description we determine the initial energy density and other thermodynamic quantities in those collisions.
\end{abstract}
\maketitle
\date{\today}

\section{Introduction}
Starting from Landau's seminal paper~\cite{Landau:1953gs} and the Hwa-Bjorken solution~\cite{Hwa:1974gn,Bjorken:1982qr}, the application of relativistic hydrodynamics for high energy physics has a long and illustrious history. Hydrodynamic calculations allow us to study the properties of the strongly interacting Quark-Gluon Plasma (QGP or sometimes denoted as sQGP). Relativistic hydrodynamic models provide a valid description of a broad range of experimental data in heavy ion collisions~\cite{Csanad:2003qa,Csanad:2004mm} as well as proton-proton and hadron-proton collisions~\cite{Agababyan:1997wd,Csorgo:2004id}. One can use relativistic hydrodynamic simulations to study the quark-hadron phase transition~\cite{Heinz:2001xi,Bass:2008rv}, elliptic flow~\cite{Huovinen:2001cy,Kolb:2003dz}, viscosity~\cite{Song:2010mg}, vorticity~\cite{Pang:2016igs} and many other aspects of high energy heavy ion physics. Results from hydrodynamic calculations for p+p, Cu+Cu, Au+Au, Pb+Pb collisions at RHIC and LHC have provided a comprehensive comparison between experimental data and models, such as the model on 1+1 dimensional hydrodynamical description~\cite{Jiang:2015apa}, the QGP medium response to jet quenching~\cite{Chen:2016owv}. Hydrodynamics has also been used to provide a basic estimate for the initial energy density using the Bjorken estimate~\cite{Bjorken:1982qr}, see for example Refs~\cite{Abelev:2008ab,Adare:2015bua,Adam:2015kda}. However, such estimates of the initial energy density have to take into account the longitudinal acceleration.

Recently, based on the successful Buda-Lund hydrodynamic model~\cite{Csorgo:1995bi,Csanad:2003qa}, a class of analytic, exact, 1+$d$ dimensional, accelerating hydrodynamic solutions~\cite{Csorgo:2006ax} has been presented. Applying these solutions to describe rapidity density $dN/dy$ results, one can extract the flow element's longitudinal acceleration parameter $\lambda$ and obtain an improved initial energy density estimation of the QGP~\cite{Nagy:2007xn}. Such a study for accelerating hydrodynamics has also been used to study p+p collisions at $\sqrt{s}=7$ and 8 TeV from CMS and TOTEM~\cite{Csanad:2013lba,Csanad:2016add}. In these cases, an advanced estimate of the initial energy density was provided, yielding approximately $0.6$ GeV/fm$^{3}$, for the average multiplicity.

In this paper, we apply the previously mentioned class of acceleration hydrodynamic solutions of perfect relativistic hydrodynamics~\cite{Nagy:2007xn} and combine it with the Buda-Lund model~\cite{Csorgo:1995bi,Csanad:2003qa} to estimate the initial energy density in various collision systems and center of mass energies at RHIC and LHC. We provide a detailed picture of charged particle pseudorapidity distributions ($dN/d\eta$), applicable for the aforementioned collision systems. Based on a hydrodynamic model describing acceleration, and the experimental data, we extract acceleration parameters ($\lambda$) for these different systems. The extracted results show that with the increase of center of mass energy $\sqrt{s_{_{NN}}}$, the longitudinal acceleration $\lambda$ decreases, while at the same center of mass energy, it increases with the multiplicity or centrality. We also find that the $\lambda$ change with multiplicity is less pronounced in case of $\sqrt{s_{_{NN}}}=2.76$ TeV PbPb collisions. These features of $\lambda$ may offer insights to study the longitudinal acceleration also in viscous hydrodynamics. Based on the obtained acceleration values, we estimate the initial energy density, temperature, and pressure for various collision systems at RHIC and LHC.

The organization of the paper is as follows. In Section~\ref{s:dndeta} we describe the hydrodynamic solutions and calculate
pseudorapidity densities. In Section~\ref{s:estimate} we detail the advanced initial energy density estimate. In Section~\ref{s:cent} the centrality dependent analysis is discussed. In Section~\ref{s:exp} the accelerating hydrodynamic solution is applied to describe RHIC and LHC data for various systems, and the initial energy density, pressure and temperature estimates are presented. Finally, in Section~\ref{s:concl} summary and conclusions are given.

\section{Pseudorapidity distributions from hydrodynamics}\label{s:dndeta}
In this section, we discuss how pseudorapidity densities are obtained from perfect fluid hydrodynamics. We adopt the following notations in this paper: $g^{\mu\nu}$ is the metric tensor, $u^{\mu}$ is the four-velocity, $n$ is the density of a conserved charge, $\epsilon$ is the energy density, $p$ is the pressure and $T$ is the temperature. We also utilize the Equation of State (EoS) $\epsilon = \kappa p$, where $\kappa$ may depend on the temperature $T$. In the case of a perfect hydrodynamics, the energy-momentum tensor in the Landau frame is
\begin{equation}
T^{\mu\nu}=(\epsilon+p)u^{\mu}u^{\nu}-pg^{\mu\nu}.
\label{1}
\end{equation}
The local continuity and energy-momentum conservation laws are given as
\begin{equation}
\partial_{\nu}(nu^{\nu})=0,\qquad\qquad  \partial_{\nu}T^{\mu\nu}=0.
\label{2}
\end{equation}
By projecting the above hydrodynamic conservation equations into components orthogonal and parallel to $u^{\mu}$, one obtains the relativistic Euler equation, the energy conservation equation and the continuity equation (for one conserved charge) :
\begin{align}
\frac{\omega}{1-v^{2}}\frac{d\vec{v}}{dt}&=-(\nabla p+\vec{v} \frac{\partial p}{\partial t}),\label{3}\\
\frac{1}{\omega}\frac{d\epsilon}{dt}&=-(\nabla \vec{v})-\frac{1}{1-v^{2}}\frac{d}{dt}\frac{v^{2}}{2},\label{4}\\
\frac{d}{dt}\ln\frac{n}{\sqrt{1-v^{2}}}&=-(\nabla \vec{v}).\label{5}
\end{align}
We use the well-known Rindler coordinates $\tau$ and $\eta$ as independent variables here, with  $\tau=\sqrt{t^{2}-r^{2}}$ being the coordinate proper-time and $\eta_{_{\rm S}}=0.5\log((t+r)/(t-r))$ the space-time rapidity. For simplicity, we assume a temperature independent EoS, $\kappa\neq\kappa(T)$, and we search for spherically symmetric solutions in 1+d dimensions, $x^\mu=(t,r_1,\dots,r_d)$ and $r=\sqrt{\Sigma_i r_i^2}$. Then we parametrize the velocity with $\Omega(\tau,\eta_{_{\rm S}})$ as $v=\tanh\Omega(\tau,\eta_{_{\rm S}})$, where $\Omega$ is the rapidity of the flow element. With calculations shown in detail in Ref.~\cite{Csorgo:2006ax}, one obtains exact analytic solutions for the above presented hydrodynamic equations, which depends on the acceleration parameter $\lambda$~\cite{Nagy:2007xn}. Table~\ref{t:solutions} presents the parameters of five different classes of accelerating hydrodynamic solutions, valid for different values of acceleration parameter $\lambda$, number of spatial dimensions $d$, EoS parameter $\kappa$ and auxiliary rapidity parameter $\phi_\lambda$.

\begin{table}
\begin{center}
\begin{tabular}{l c c c c}
\hline\hline
case     & $\qquad$ $\lambda$      &   $\qquad$ $d$                     &  $\qquad$ $\kappa$             & $\qquad$ $\phi_{\lambda}$\\\hline
(a)        &  $\qquad$ 2                  &   $\qquad$ $\in \mathbb {R}$  &  $\qquad$ d                        &  $\qquad$ 0\\
(b)        &  $\qquad$ $\frac{1}{2}$  &   $\qquad$ $\in \mathbb {R}$  &   $\qquad$1                        &  $\qquad$ $\frac{k+1}{k}$ \\
(c)        &  $\qquad$ $\frac{3}{2}$  &   $\qquad$ $\in \mathbb {R}$  &   $\qquad$ $\frac{4d-1}{3}$   & $\qquad$ $\frac{k+1}{k}$ \\
(d)        &  $\qquad$ 1                 &   $\qquad$ $\in \mathbb {R}$  &   $\qquad$ $\in \mathbb {R}$ &  $\qquad$ 0 \\
(e)        &  $\qquad$ $\in \mathbb {R}$ &   $\qquad$1                   &   $\qquad$1                         &  $\qquad$ 0 \\
\hline\hline
\end{tabular}
\end{center}
\caption{The five different cases of solutions, from Refs~\cite{Csorgo:2006ax,Nagy:2007xn}.}\label{t:solutions}
\end{table}

In all cases, the velocity field and the pressure takes the following form
\begin{align}
v&=\tanh \lambda\eta_{_{\rm S}},\label{6}\\
p&=p_{0}\left(\frac{\tau_{0}}{\tau}\right)^{\lambda {\rm d}\frac{\kappa+1}{\kappa}}\left(\cosh(\frac{\eta_{_{\rm S}}}{2})\right)^{-({\rm d}-1)\phi_{\lambda}},
\label{7}
\end{align}
where $p_{0}$ and $\tau_{0}$ define the initial values for pressure and thermalization time. The properties of these accelerating, exact solutions are detailed in Refs.~\cite{Nagy:2007xn}.

Combining accelerating hydrodynamics and the Cooper-Frye flux term~\cite{Cooper:1974mv} in the Boltzmann approximation, one can obtain momentum distributions as a function of four-momentum components $(E,p_x,p_y,p_z)$, three-momentum length $p=\sqrt{\Sigma_i p_i^2}$, transverse momentum $p_T=\sqrt{p_x^2+p_y^2}$, pseudorapidity $\eta=0.5\log((p+p_z)/(p-p_z))$ or rapidity $y=0.5\log((E+p_z)/(E-p_z))$. The pseudorapidity distribution $dN/d\eta$~\cite{Csorgo:1995bi,Csorgo:2006ax} in terms of the rapidity distribution $dN/dy$ can be given as as~\cite{Csorgo:2006ax,Csanad:2013lba,Csanad:2016add}
\begin{equation}
\frac{dN}{d\eta}\simeq\left.\frac{\bar{p}_{T}}{\bar{E}}\frac{dN}{dy}\right|_{y=\eta}
=\left.\frac{\bar{p}_{T}\cosh \eta}{\sqrt{m^{2}+\bar{p}^{2}_{T}\cosh^{2}\eta}}\frac{dN}{dy}\right|_{y=\eta},
\label{8}
\end{equation}
where $m$ is the average mass of the charged particles, $\bar{p}_{T}$ is the mean transverse momentum, and the Jacobian connecting rapidity and pseudorapidity has been utilized~\cite{Csorgo:2006ax}. Based on the Buda-Lund hydrodynamic model~\cite{Csorgo:1995bi,Csanad:2003qa}, in the region of $p_{T}$ $<$ 2 GeV, the relation between mean transverse momentum and the the effective temperature $T_{\rm eff}$ at a given rapidity $y$ can be written as
\begin{equation}
\bar{p}_{T}=\frac{T_{\rm eff}}{1+\frac{\sigma^{2}}{2}(y-y_{\rm mid})^{2}},
\label{9}
\end{equation}
where $\sigma$ parameterizes the effective temperature gradient, and $y_{\rm mid}$ is the central rapidity. The rapidity distribution for $y_{\rm mid}=0$, as calculated in Refs~\cite{Csorgo:2006ax,Nagy:2007xn,Csanad:2013lba,Csanad:2016add}, is then:
\begin{equation}
\frac{dN}{dy}\simeq N_{0}
\cosh^{-\frac{\alpha}{2}-1}\left(\frac{y}{\alpha}\right)\exp\left[-\frac{m}{T_{f}}\cosh^{\alpha}\left(\frac{y}{\alpha}\right)\right],
\label{10}
\end{equation}
where $\alpha=\frac{2\lambda-1}{\lambda-1}$ is a derived acceleration parameter, $T_{f}$ is the freeze-out temperature with typical values around $90-170$ MeV, $N_{0}=\sqrt{\frac{2\pi mT^{3}_{f}}{\lambda(2\lambda-1)}}\frac{S_{\perp}m\tau_{f}}{2\pi \hbar}$ is a normalization constant, with $S_{\perp}$ being the transverse cross section of the fluid. In Section~\ref{s:exp}, the above calculated $dN_{\rm ch}/d\eta$ approximation is used to determine the acceleration parameter $\lambda$ for the various collision systems and energies.

\section{The energy density estimate}\label{s:estimate}

An important consequence of the previously discussed result for the pesudorapidity density is that it allows for an improved initial energy density estimate. The accelerationless Hwa-Bjorken-flow yields an initial energy density estimate $\epsilon_{\rm Bj}$, the Bjorken-estimate~\cite{Bjorken:1982qr}. In this case, a thin transverse overlap area of the two nucleus at midrapidity at the thermalization time ($\tau=\tau_{0}$) is considered, and the energy density is then estimated from the finial state~\cite{Adare:2015bua}. The Bjorken-estimate can thus be expressed at midrapidity as
\begin{equation}
\epsilon_{\rm Bj}=\frac{1}{S_{\perp}\tau_{0}}\frac{d\langle E\rangle}{d\eta}
=\frac{\langle E\rangle}{S_{\perp}\tau_{0}}\frac{dN}{dy},
\label{11}
\end{equation}
where $S_{\perp}$ can be understood as the transverse overlap area of the colliding nuclei, and $\tau_{0}$ is the proper-time of thermalization, which was estimated by Bjorken to be $\tau_{0}=1\rm{fm}/c$. For the most central collisions of identical nucleii, the transverse area can be approximated as $S_{\perp}=\pi R^{2}$, with $R$ being the nuclear radius, $R=1.18A^{1/3}$ fm. For non-central collisions, this can be calculated via Glauber calculations~\cite{Abbas:2013taa,Adam:2016thv}, as we will discuss subsequently. The volume element of this system is $dV=(R^{2}\pi)\tau d\eta_{_{\rm S}}$, where $d\eta_{_{\rm S}}$ is the space-time rapidity element corresponding to the volume. The energy content in this volume is $dE=\langle E \rangle dN$. One may then utilize experimental $dE/dy$ results e.g. from Refs.~\cite{Adare:2015bua,Adam:2015kda} to estimate the Bjorken energy density. Alternatively, average transverse mass or transverse momentum may also be used, via $\langle m_{t}\rangle=\sqrt{\langle p_{T} \rangle^{2}+m^{2}}$, determined from $\pi^{\pm}$, $K^{\pm}$, $p$ and $\bar{p}$ transverse momentum distributions at midrapidity presented~\cite{Abelev:2008ab}.

For accelerationless, boost-invariant Hwa-Bjorken flow, the initial and final space-time rapidities coincide with the momentum rapidity: $\eta_{_{\rm S,0}}=\eta_{_{\rm S,f}}=y$. However, in case of longitudinally accelerating flow, one has to apply a correction to take into account the acceleration effects on the energy density. Given an acceleration parameter $\lambda\in\mathbb{R}$, the maximal particle production occurs at $y\neq \eta_{_{\rm S,f}}$, which yields a correction factor of $\frac{\partial y}{\partial \eta_{_{\rm S,f}}}=(2\lambda-1)$. The volume element is also changed by a factor of $\frac{\partial \eta_{_{\rm S,f}}}{\partial\eta_{_{\rm S,0}}}=(\frac{\tau_{f}}{\tau_{0}})^{\lambda-1}$, see Ref.~\cite{Nagy:2007xn} for details. The initial energy density that corresponds to a given final state is also dependent on the EoS parameter $\kappa$. A conjecture that is consistent with known exact results for the $\lambda=1$ or the $\kappa=1$ case, and also consistent with numerical results, was put forward in Ref.~\cite{Csorgo:2006ax}. This conjectured initial energy density is given by a corrected estimate $\epsilon_{\rm corr}$ as~\cite{Csanad:2016add}
\begin{equation}
\epsilon_{\rm corr} = (2\lambda-1)\left(\frac{\tau_{f}}{\tau_{0}}\right)^{\lambda-1}\left(\frac{\tau_{f}}{\tau_{0}}\right)^{(\lambda-1)(1-\frac{1}{\kappa})}\epsilon_{\rm Bj}.
\label{12}
\end{equation}
This advanced estimate is based on the acceleration parameter $\lambda$ (determined from pseudorapidity density measurements), initial proper time $\tau_{0}$, freeze-out proper time $\tau_{f}$. In Refs.~\cite{Csorgo:2006ax,Nagy:2007xn}, the accelerating hydrodynamic model was fitted to rapidity distributions measured by BRAHMS in 200 GeV $0-5\%$ centrality Au+Au collisions, and assuming $\tau_{f}/\tau_{0}=8\pm2$, an advanced estimate of $\epsilon_{\rm corr}\simeq(10.0\pm 0.5)$ GeV/$\rm{fm^{3}}$ was obtained for $\kappa=1$, while for realistic $\kappa\approx 7-10$ values $\epsilon_{\rm corr}\simeq (14.5\pm0.5)$ GeV/fm$^3$ was obtained in Ref.~\cite{Csorgo:2006ax}.

\section{Centrality dependent energy density estimate}\label{s:cent}

In case of non-central collisions, several properties used in estimating the initial energy density are different from the most central collisions. With the increase of impact parameter, the volume of the created fireball decreases. Based on the experimental data~\cite{Adam:2015kda,Adare:2015bua}, one can use the Glauber Monte Carlo model of Ref~\cite{Alver:2008zza} to obtain $S_\perp$, the transverse overlap of the two colliding nuclei. Also, it is necessary to discuss the change of initial proper time $\tau_{0}$ for different collision energies and centralities. It is reasonable to assume that $\tau_0$ is anticorrelated with $\sqrt{s_{_{NN}}}$, and is not necessarily correlated with the centrality at a fixed $\sqrt{s_{_{NN}}}$. The initial central temperature $T_{0}$ is inversely related to the initialization time $\tau_{0}$, and $T_0$ values of 500 MeV and 650 MeV for RHIC and LHC~\cite{Kolb:2003dz} may be used then to estimate $\tau_0$. This relationship results in a rough estimate for the value of proper time $\tau_{0}$. Note however, that $\tau_{0}=0.6$ fm/c was given in Ref~\cite{Kolb:2003dz} for 130 GeV Au+Au collisions. In our model, freeze-out happens on a hypersurface pseudo-orthogonal to the four-velocity field when the temperature at $\eta = 0$ reaches a given $T_{f}$ value~\cite{Nagy:2007xn}. With considerations of initial equilibration time $\tau_{0}$ and freeze-out condition, the $\tau_{f}/\tau_{0}$ ratio is directly correlated with $\sqrt{s_{_{NN}}}$, but there is an inverse correlation between $\tau_{f}$ and the impact parameter $b$. Here for simplicity, we follow Bjorken's estimate for the initial energy density, and assume the proper time $\tau_{0}=1~$fm/c for different centrality dependence collisions as usual. When acceleration effects become important, the corrected initial energy density estimate is given in Eq.~\eqref{12}, which contains the influence of $\tau_{0}$ and $\tau_{f}$ correlations~\cite{Gyulassy:1983ub}. For different centralities and collision energies, the acceleration parameter $\lambda$, the transverse area $S_\perp$, and the ratio $\tau_{f}/\tau_{0}$ are different.  The experimentally given $dE/dy$ values may then also be utilized instead of $\langle E\rangle dN/dy$, to arrive at
\begin{equation}
\epsilon_{\rm corr} = (2\lambda-1)\left(\frac{\tau_{f}}{\tau_{0}}\right)^{\lambda-1}\left(\frac{\tau_{f}}{\tau_{0}}\right)^{(\lambda-1)(1-\frac{1}{\kappa})}\frac{1}{S_{\perp}\tau_{0}}\frac{dE}{dy}.
\label{13}
\end{equation}
The estimation of Eq.~\eqref{13} gives then modification of the initial energy density for various $\lambda$, $S_\perp$, $\tau_0$, $\tau_f$ and $dE/dy$ values, in case of a centrality dependent analysis.

\section{Analysis of proton-proton and nucleus-nucleus collisions at LHC and RHIC energies}\label{s:exp}
Detailed measurements of the charged particle pseudorapidity distribution $dN/d\eta$ at different $\sqrt{s_{_{NN}}}$ are available at RHIC~\cite{Adare:2015bua,Alver:2010ck} and at the LHC~\cite{Adam:2015kda,Adam:2016thv}. Hence one can extract the acceleration parameter of these systems. In this Section, we analyze a series of $dN/d\eta$ datasets, obtain acceleration parameter $\lambda$ and calculate the energy density correction ratio $\epsilon_{\rm corr}/\epsilon_{\rm Bj}$ (as a function of $\tau_{f}/\tau_{0}$). We then give the improved estimate of the initial energy density $\epsilon_{\rm corr}$, the initial temperature and the initial pressure with different Equations of State (different $\kappa$ values) as a function of multiplicity.

\begin{figure}
\begin{center}
\includegraphics[width=0.51\linewidth]{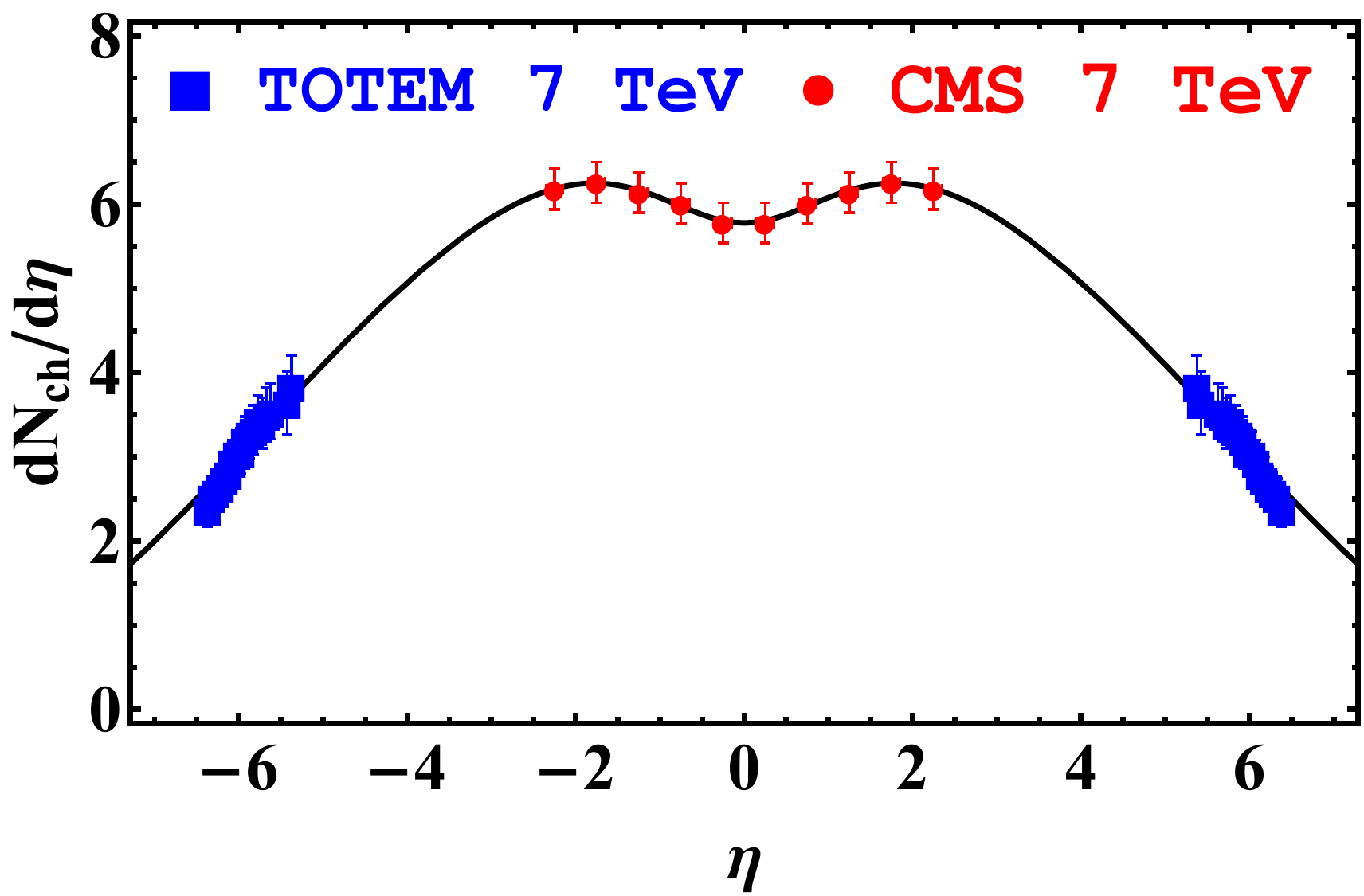}
\includegraphics[width=0.47\linewidth]{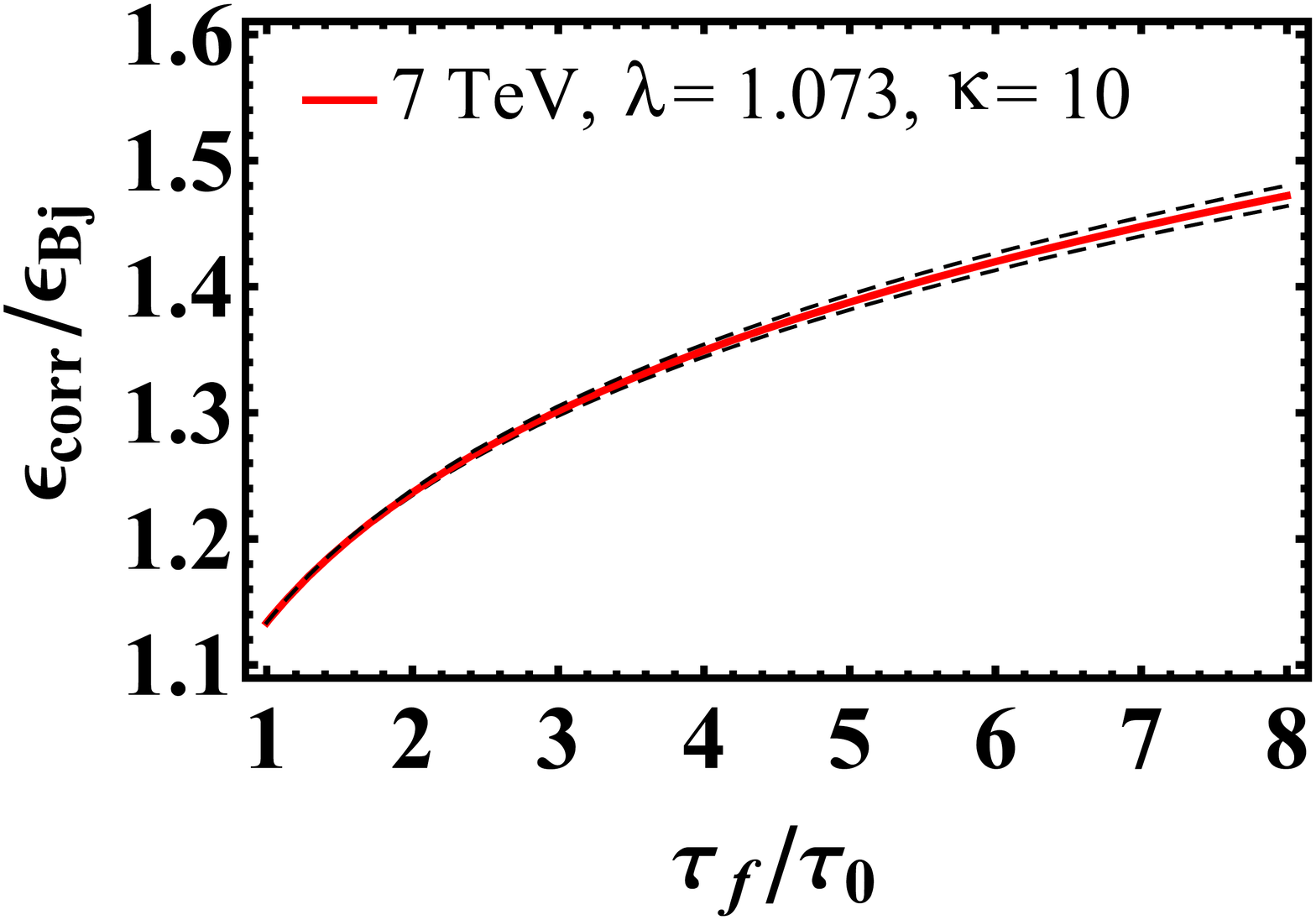}\\
\includegraphics[width=0.51\linewidth]{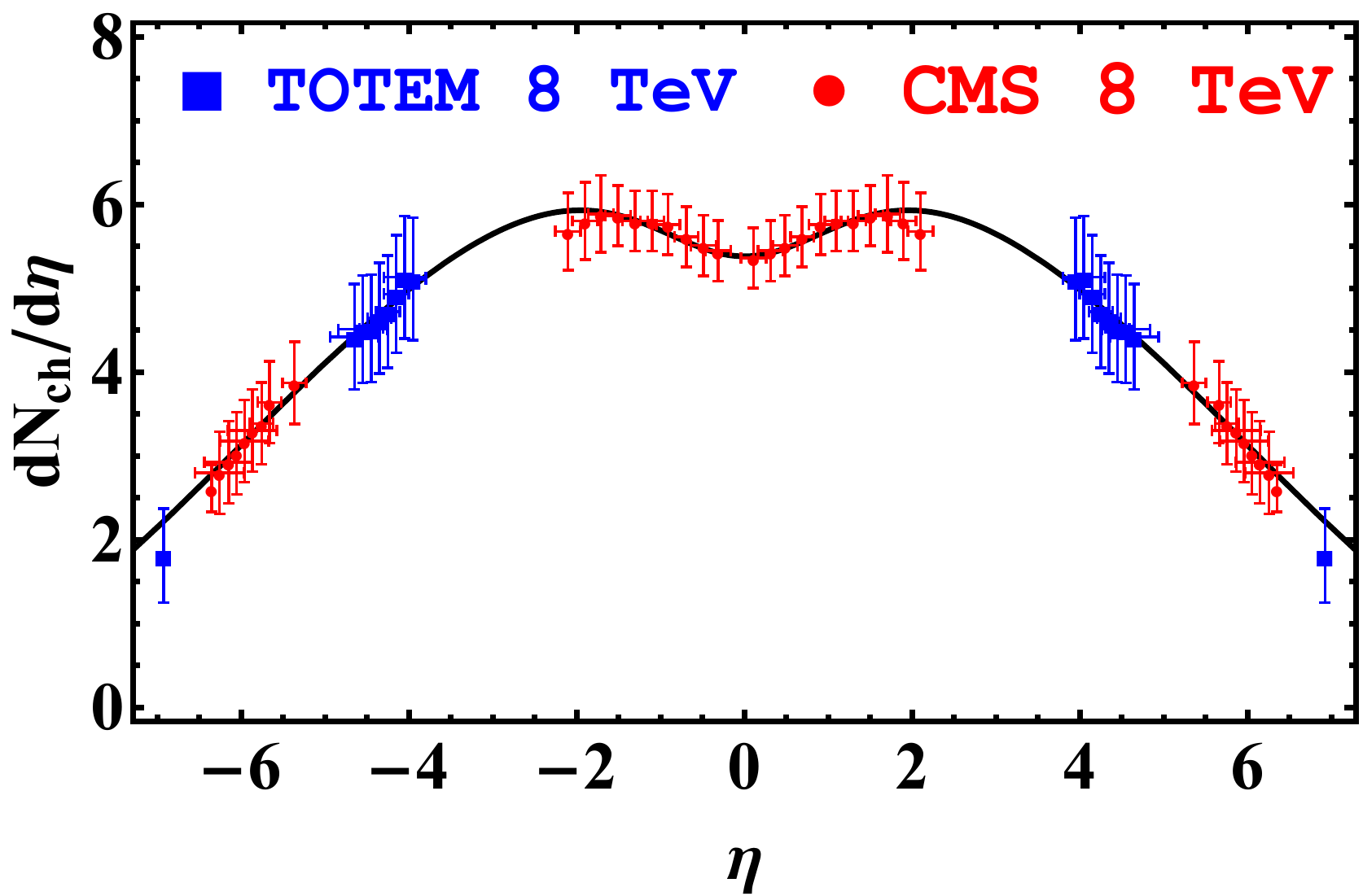}
\includegraphics[width=0.47\linewidth]{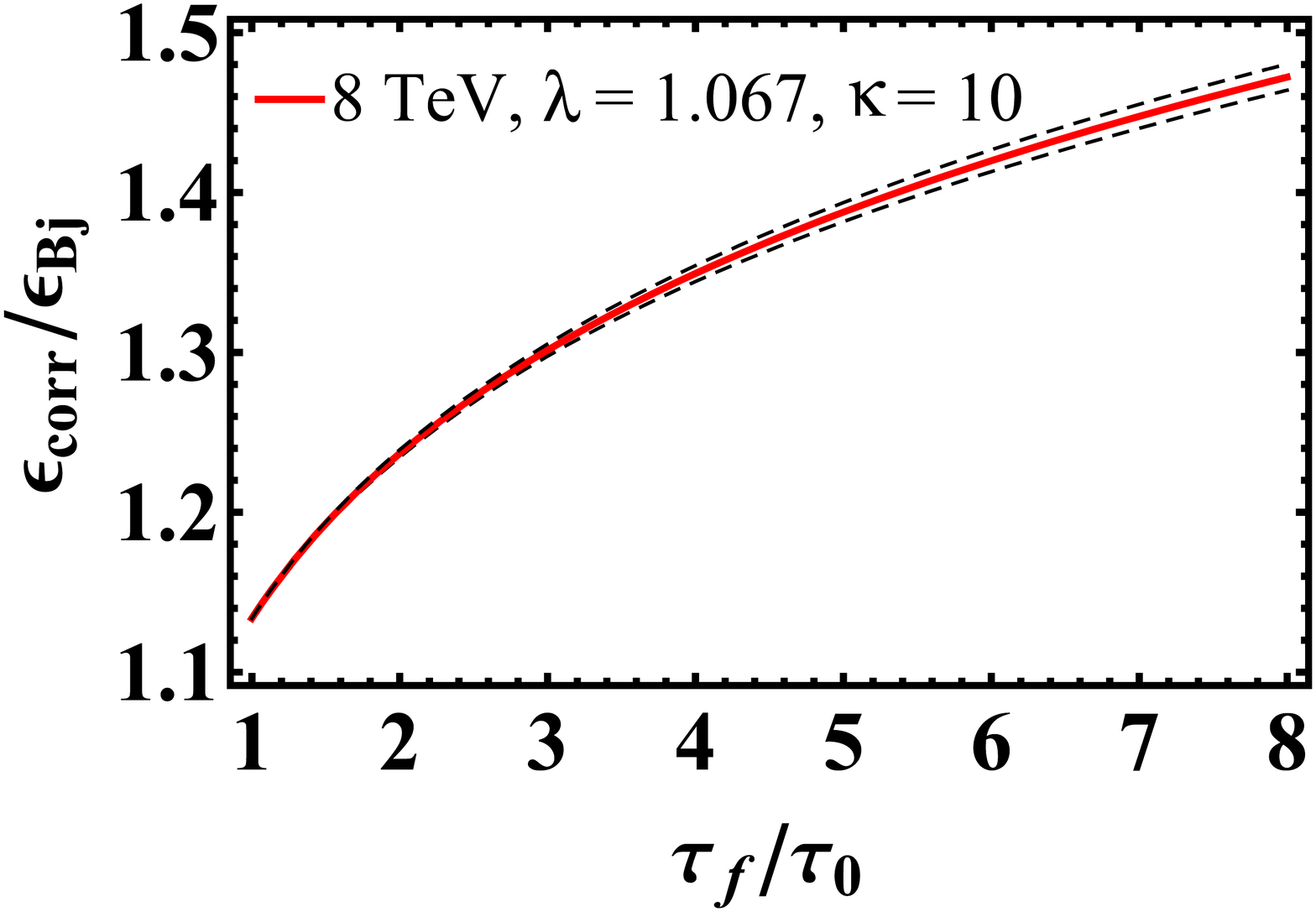}
\end{center}\vspace{-10pt}
\caption{(Color online) Left: Charged particle pseudorapidity distributions measured by CMS~\cite{Khachatryan:2010us} and TOTEM~\cite{Aspell:2012ux} at 7 TeV (first row) and 8 TeV (second row), compared to calculations from the relativistic hydrodynamic solution presented in this paper, similarly to Ref.~\cite{Csanad:2016add}. Right: The correction factor $\epsilon_{\rm corr}/\epsilon_{\rm Bj}$ is shown as a function of freeze-out time versus thermalization time ($\tau_{f}/\tau_{0}$) for the central collision (the dashed lines represent the uncertainty). }\label{f:ppfits}
\end{figure}

\begin{figure}
\begin{center}
\includegraphics[width=0.81\linewidth]{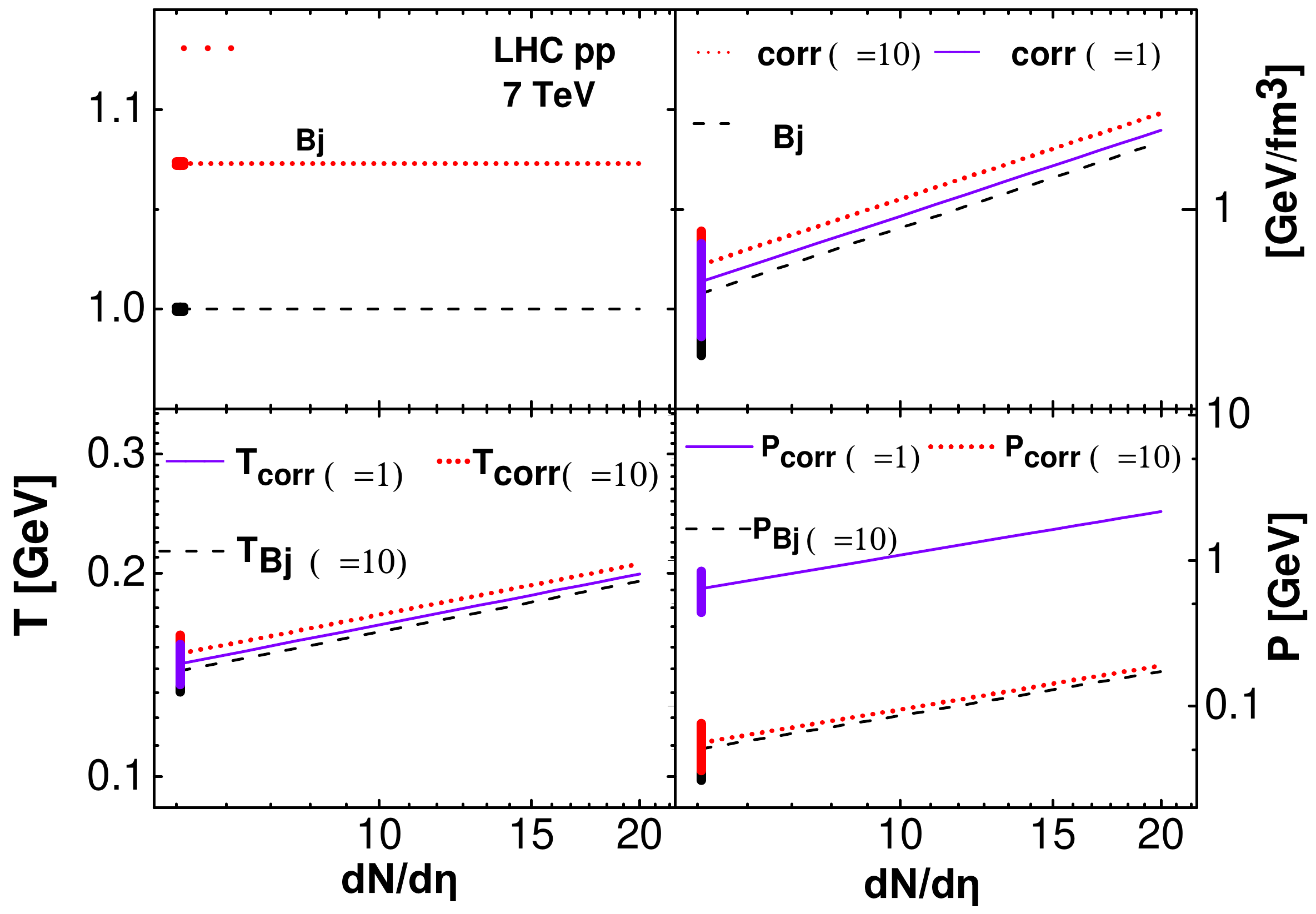}
\includegraphics[width=0.81\linewidth]{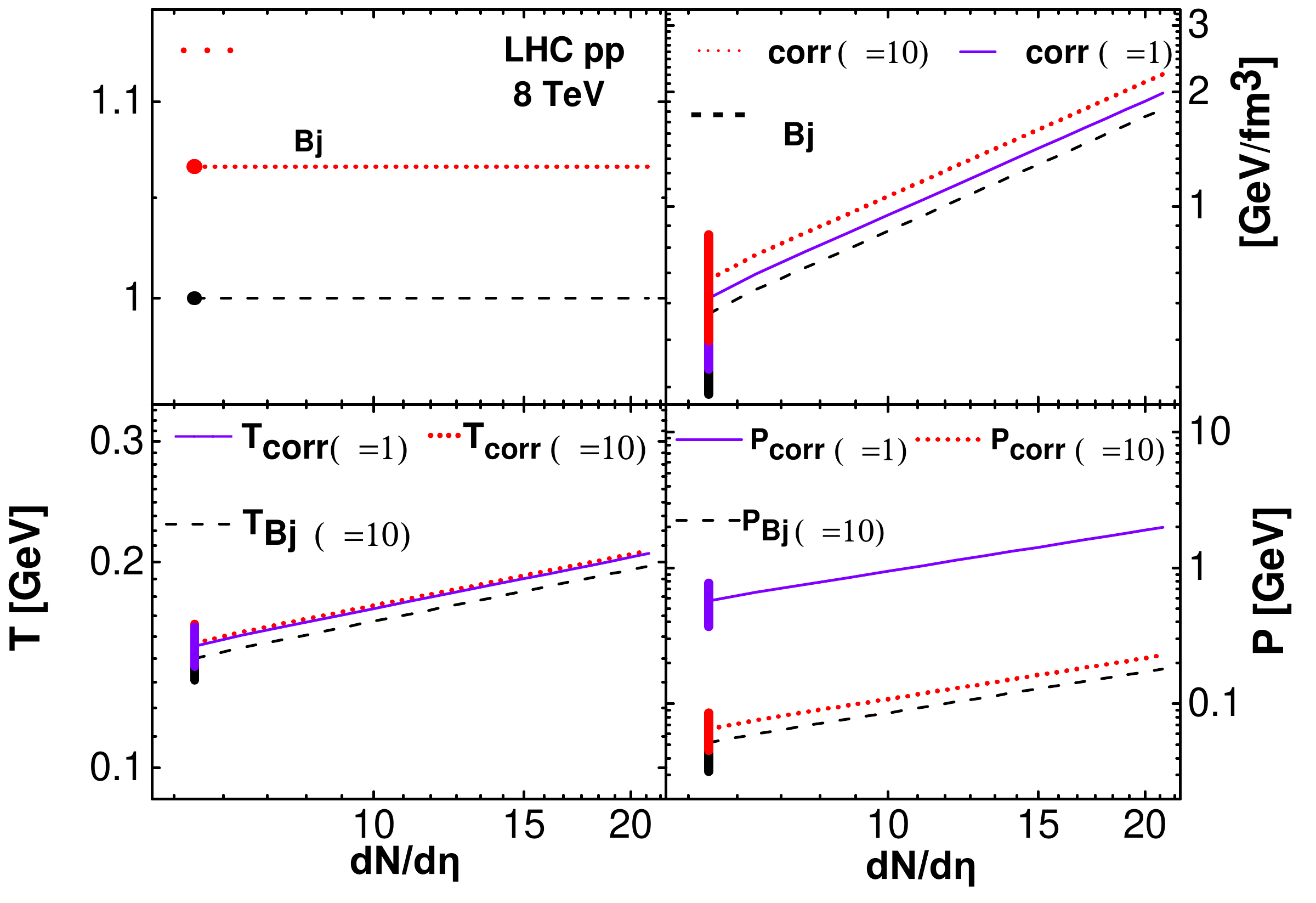}
\end{center}\vspace{-10pt}
\caption{(Color online) Acceleration parameter $\lambda$ for $\sqrt{s}=7$ and 8 TeV p+p collisions is given at the average multiplicity
of the measured pseudorapidity distributions. The calculated multiplicity dependence of the initial energy density, temperature and pressure is indicated for various EoS parameter $\kappa$ values. Systematic uncertainties are also indicated, stemming from the determination of $\tau_{f}/\tau_{0}$, $\lambda$, $dN/d\eta$, as well as from the systematic uncertainties of the data.}\label{f:epspp}
\end{figure}

\begin{table}
\begin{center}
\begin{tabular}{l c c c c c}
\hline\hline
$\sqrt{s}$ & ~$\left.\frac{dN}{d\eta}\right\vert_{\eta=\eta_{0}}$ &~$\lambda$  &~$\sigma$  &~$T_{\rm eff}[{\rm GeV}]$   &~$\chi^{2}/NDF$  \\ [0.5ex]
\hline
~$7$ TeV    &  ~5.78$\pm$0.01      &~1.073$\pm$0.001        &~0.81$\pm$0.05     &~0.18    &~0.18/22 \\
~$8$ TeV    &  ~5.36$\pm$0.02      &~1.067$\pm$0.001        &~0.86$\pm$0.13    &~0.17   &~0.30/28  \\
\hline\hline
\end{tabular}
\end{center}
\caption{Fit parameters for 7 and 8 TeV pp data, from Ref.~\cite{Csanad:2016add}. Auxiliary values of $T_{f}=0.20$ GeV, $\bar{m}=0.24$ GeV have been utilized, based on Refs.~\cite{Csorgo:1995bi,Abelev:2008ab}}\label{t:pppars}
\end{table}

\begin{table}
\begin{center}
\begin{tabular}{l c c c c  }
\hline\hline
$\sqrt{s}$  & $~\epsilon_{\rm Bj} [{\rm GeV/fm^{3}}]$ & $~\epsilon_{\rm corr} [{\rm GeV/fm^{3}}]$  & $~T_{\rm corr} [{\rm GeV}]$  &  $~P_{\rm corr} [{\rm GeV}]$ \\
\hline
~~~$7$ TeV]     &~0.51$\pm$0.01      &  0.64$\pm$0.01      &~0.16$\pm$0.001        &~~0.06$\pm$0.001            \\
~~~$8$ TeV       &~0.52$\pm$0.01     &  0.64$\pm$0.01     &~0.15$\pm$0.001        &~~0.06$\pm$0.001             \\
\hline\hline
\end{tabular}
\end{center}
\caption{Initial thermodynamic quantities obtained by the hydrodynamic fits to 7 \& 8 TeV pp data from TOTEM and CMS, using $\tau_{f}/\tau_{0}=2$.}\label{t:ppeps}
\end{table}

Before showing the results for nucleus-nucleus collisions, let us recapitulate and show the results for LHC pp collisions~\cite{Csanad:2016add}, measured by CMS~\cite{Khachatryan:2010us,Chatrchyan:2014qka} and TOTEM~\cite{Aspell:2012ux,Antchev:2014lez} at $\sqrt{s}=7$ TeV and 8 TeV. From the fits to CMS and TOTEM $dN/d\eta$ data, shown in the left panel of Fig.~\ref{f:ppfits} and detailed in Table~\ref{t:pppars}, longitudinal acceleration parameters $\lambda=1.073\pm0.001$ ($\sqrt{s}=7$ TeV) and $\lambda=1.067\pm0.001$ ($\sqrt{s}=8$ TeV) are obtained. These yield an estimate for the initial energy density $\epsilon_{\rm corr}=0.640$ GeV/fm$^{3}$ at 7 TeV, and $\epsilon_{\rm corr}=0.644$ GeV/fm$^{3}$ for 8 TeV. Let us note, that $\epsilon_{\rm corr}$ as well as $\epsilon_{\rm Bj}$ is directly proportional to the charged particle multiplicity $dN/d\eta$, so in large multiplicity event classes, $\epsilon_{\rm Bj}\geq 1$ GeV/fm$^3$ initial energy density can be reached, as illustrated in Fig.~\ref{f:epspp}. We may also estimate the initial temperature and pressure, based on the $\epsilon \propto T^{4}$ relationship and the EoS relationship $\epsilon=\kappa p$~\cite{Huovinen:2009yb,Bazavov:2009zn}. We may use $\kappa=10$, i.e. a speed of sound of $c_s\approx 0.32$~\cite{Lacey:2006bc,Borsanyi:2010cj,Csanad:2011jq}. Values are given in Table~\ref{t:ppeps}, and for more details, see Ref.~\cite{Csanad:2016add}.

\begin{figure}
\vspace{-10pt}
\begin{center}
\includegraphics[width=0.47\linewidth]{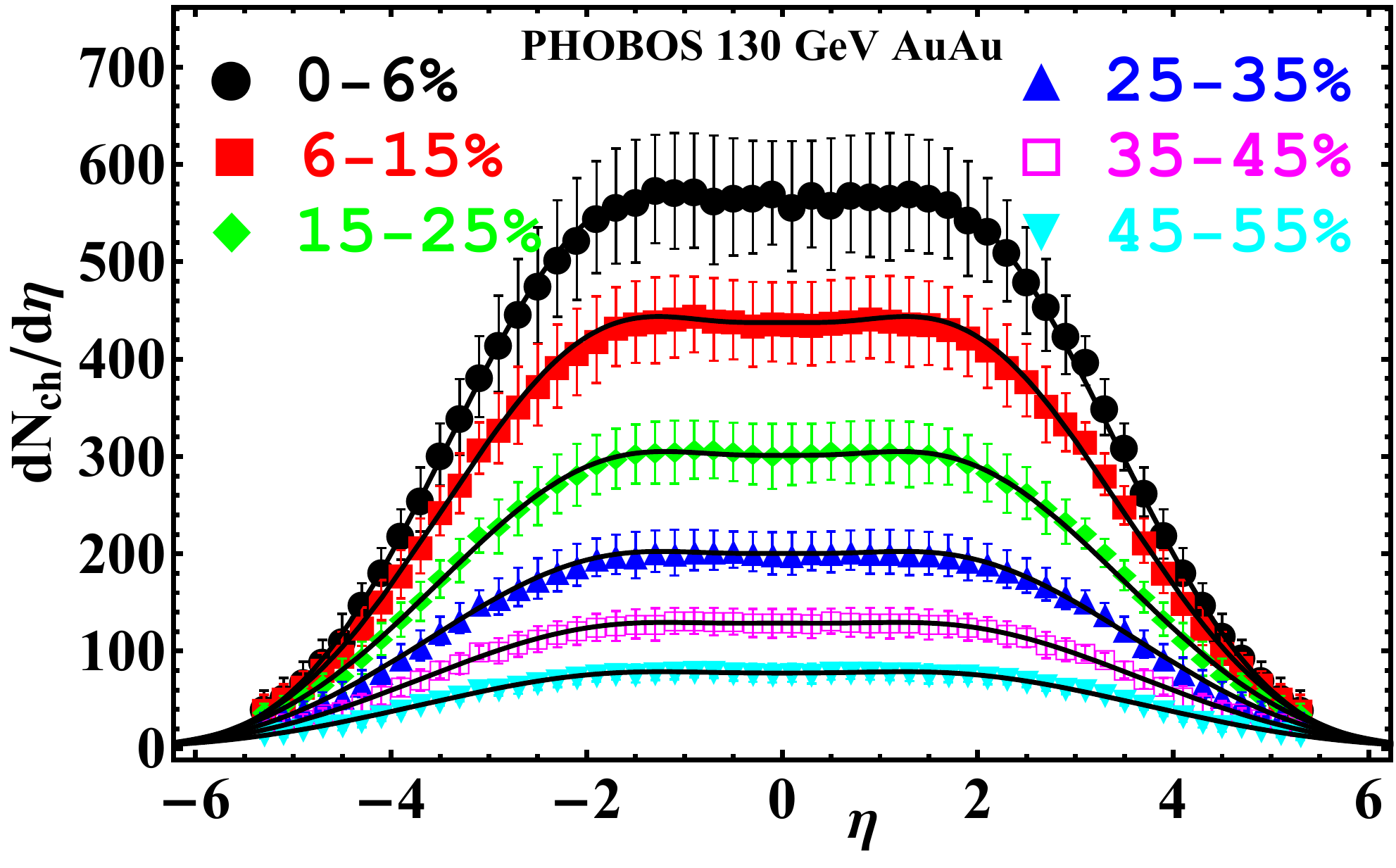}
\includegraphics[width=0.45\linewidth]{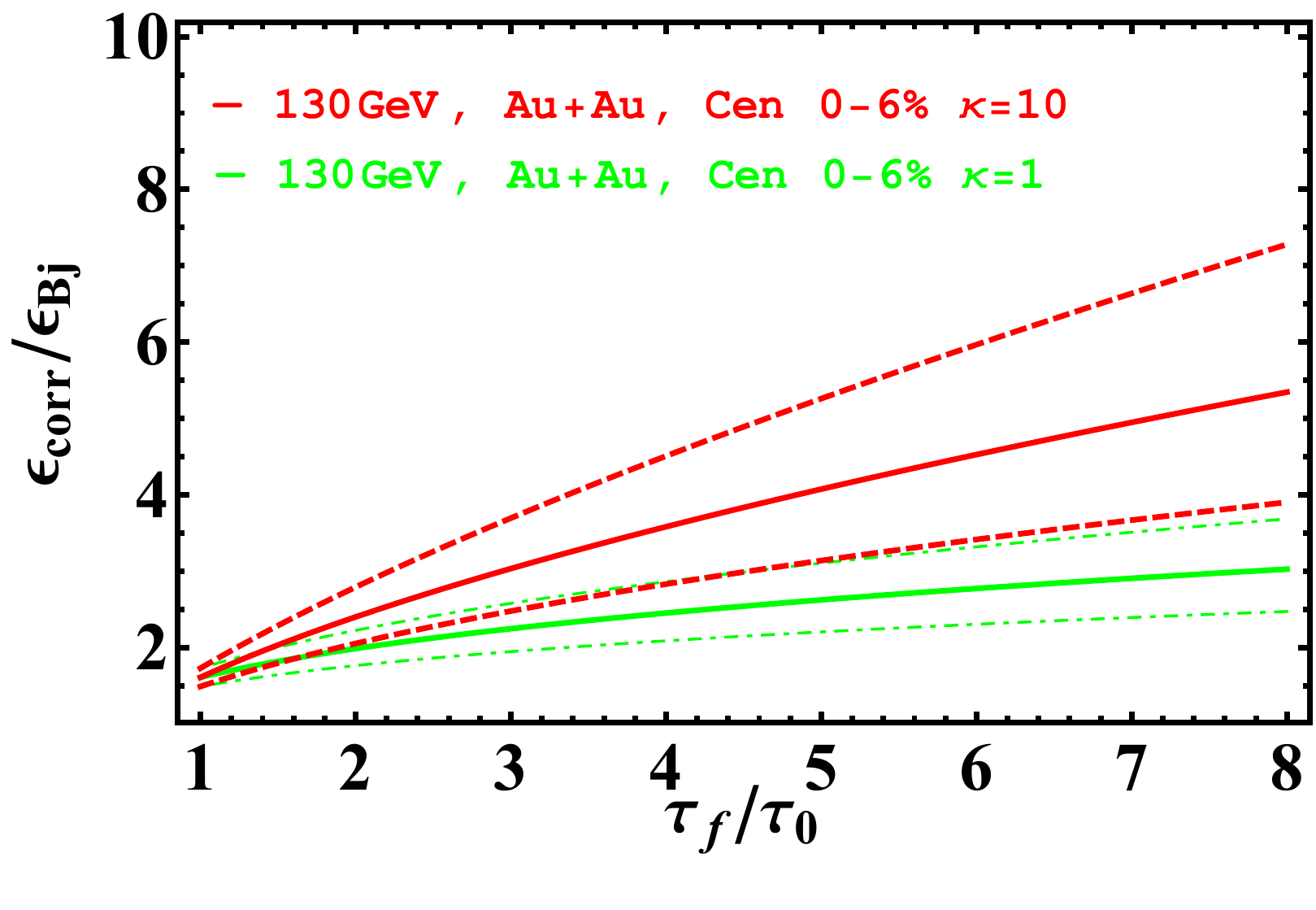}
\includegraphics[width=0.47\linewidth]{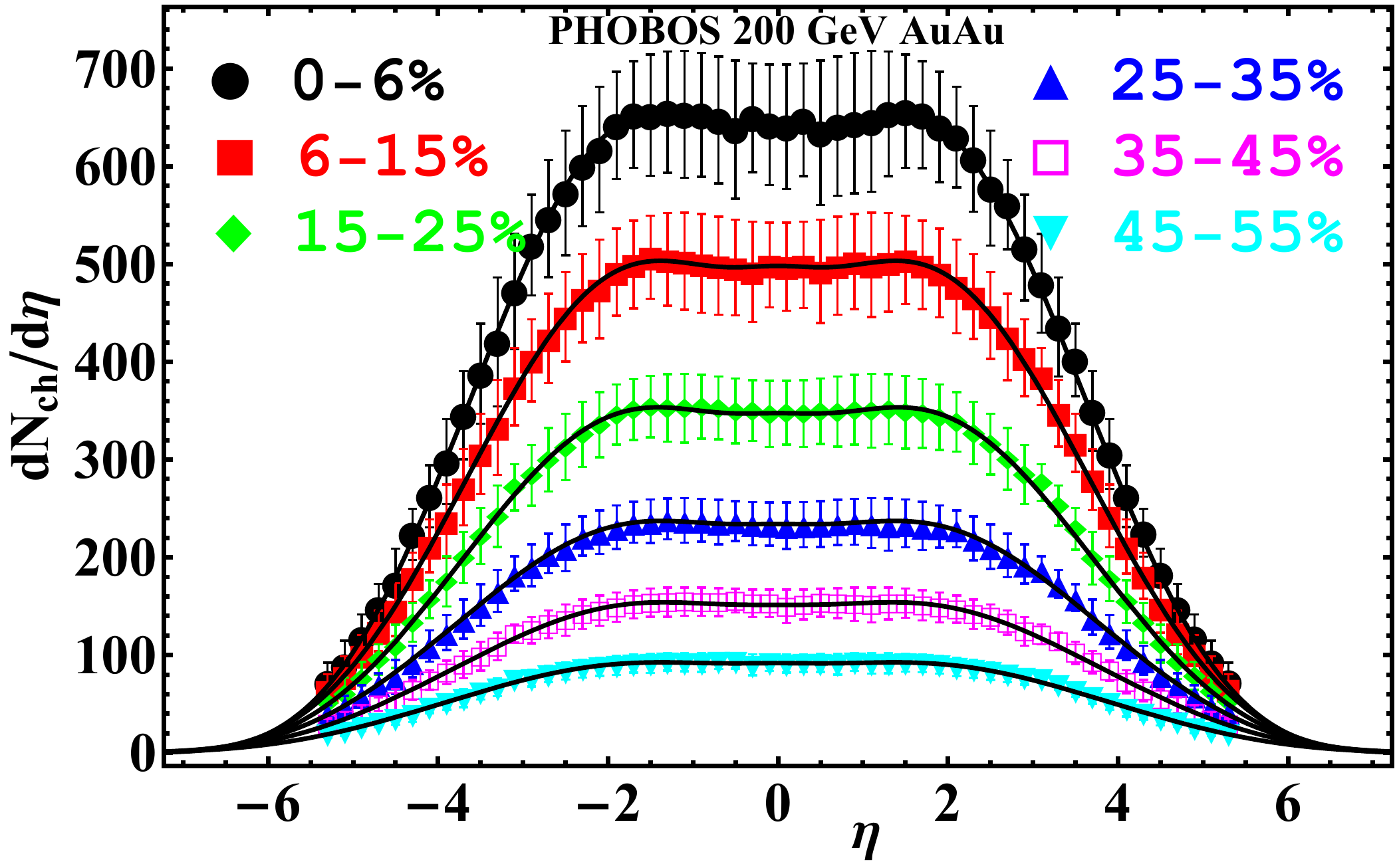}
\includegraphics[width=0.45\linewidth]{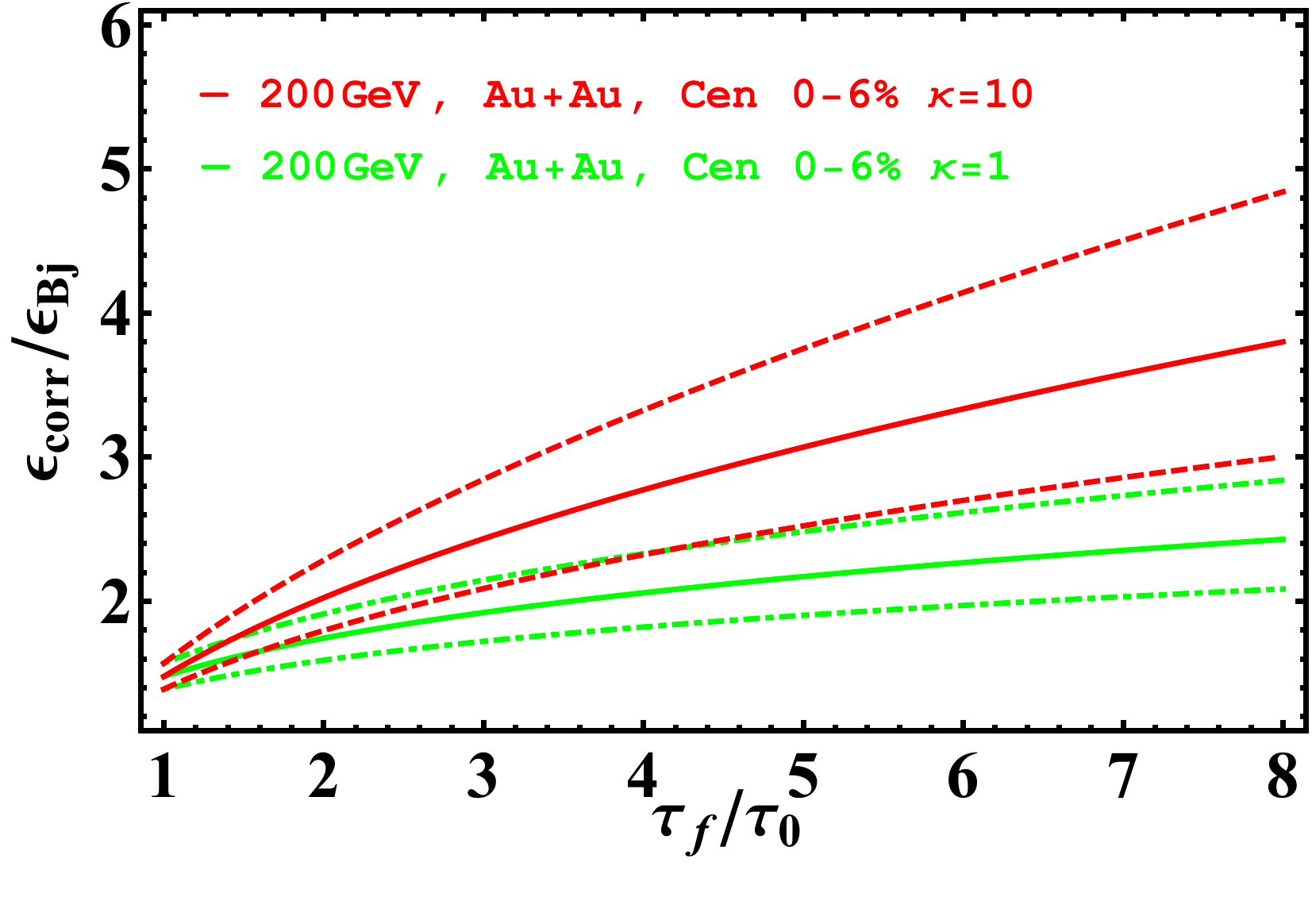}
\includegraphics[width=0.47\linewidth]{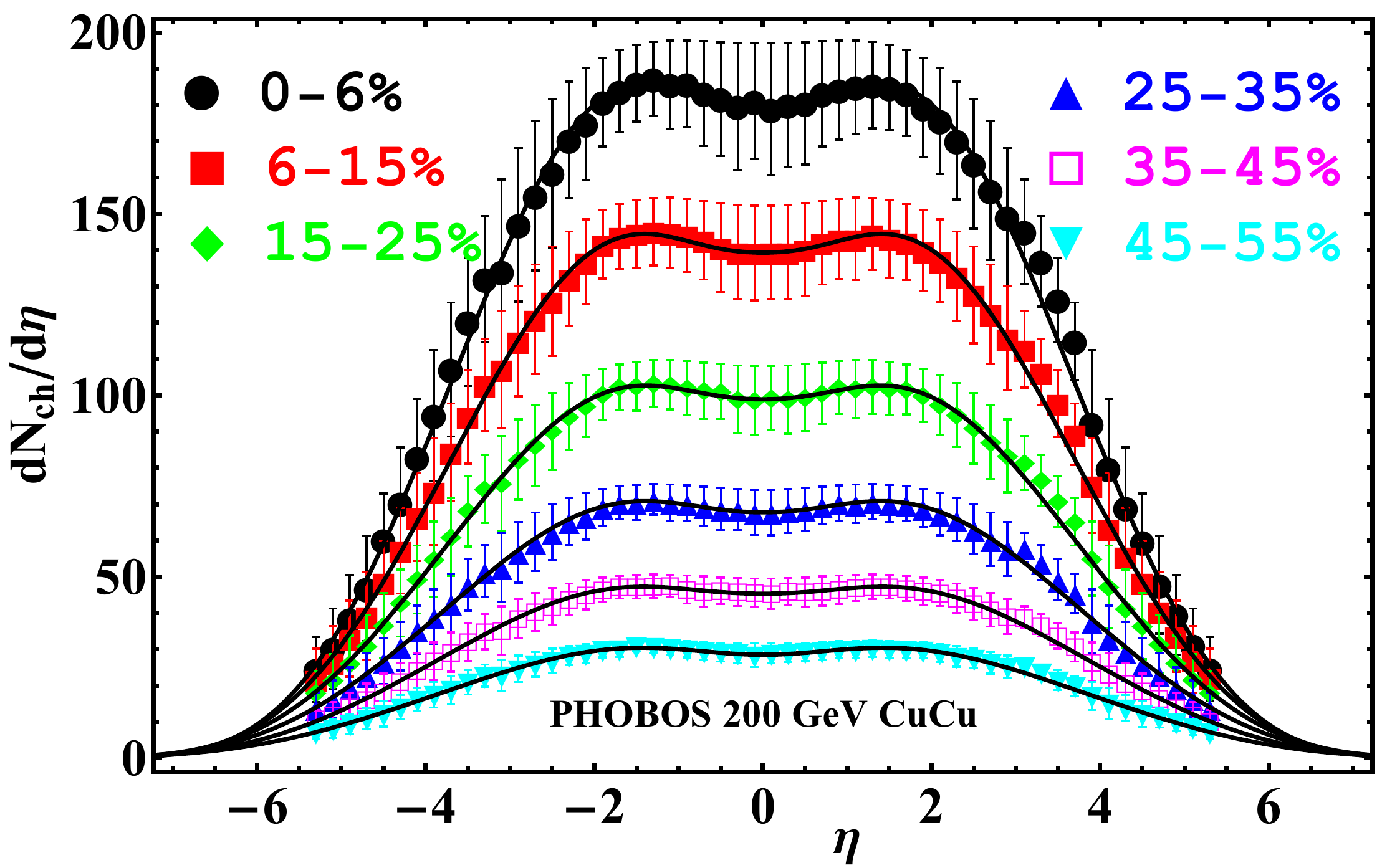}
\includegraphics[width=0.45\linewidth]{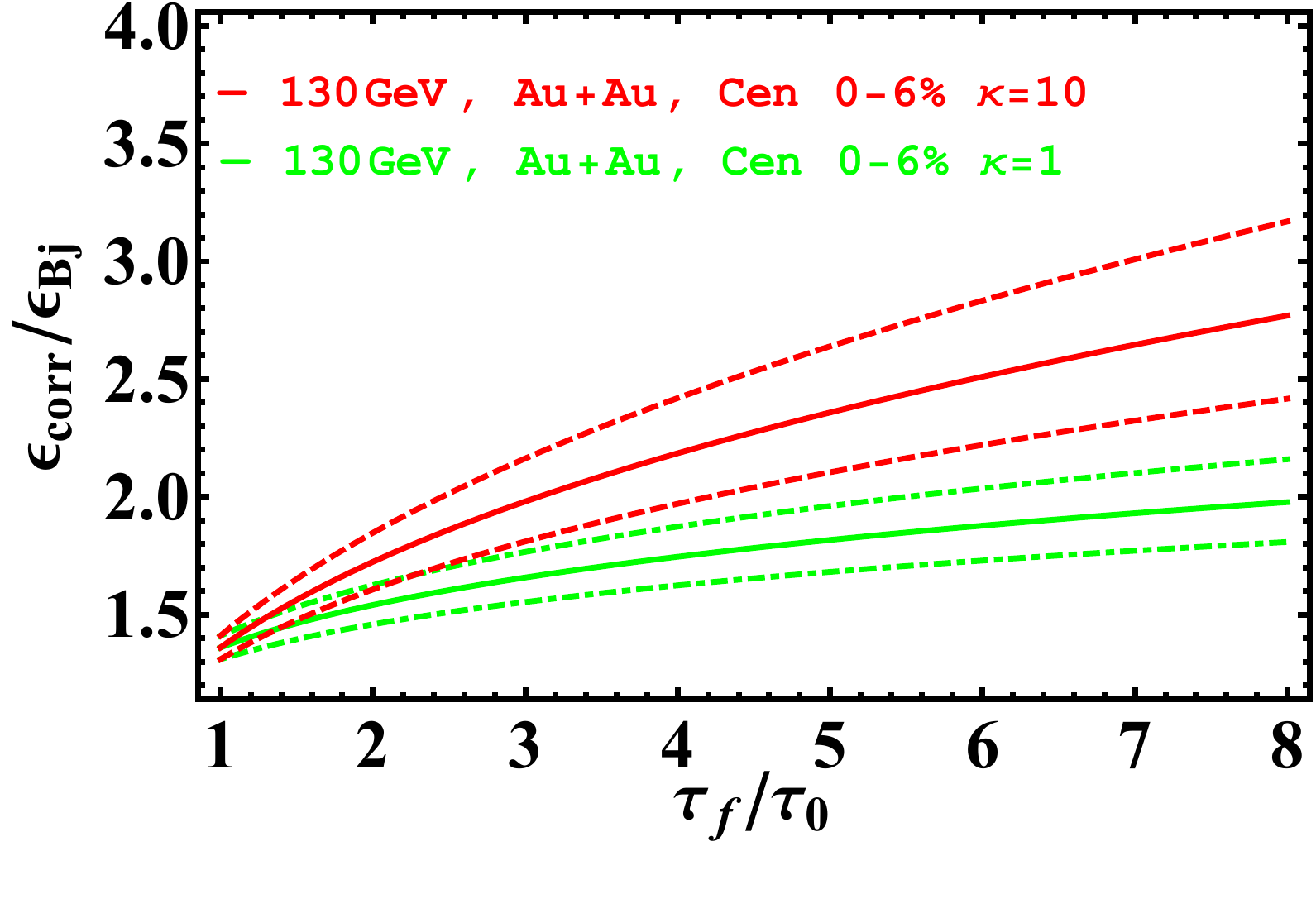}
\includegraphics[width=0.47\linewidth]{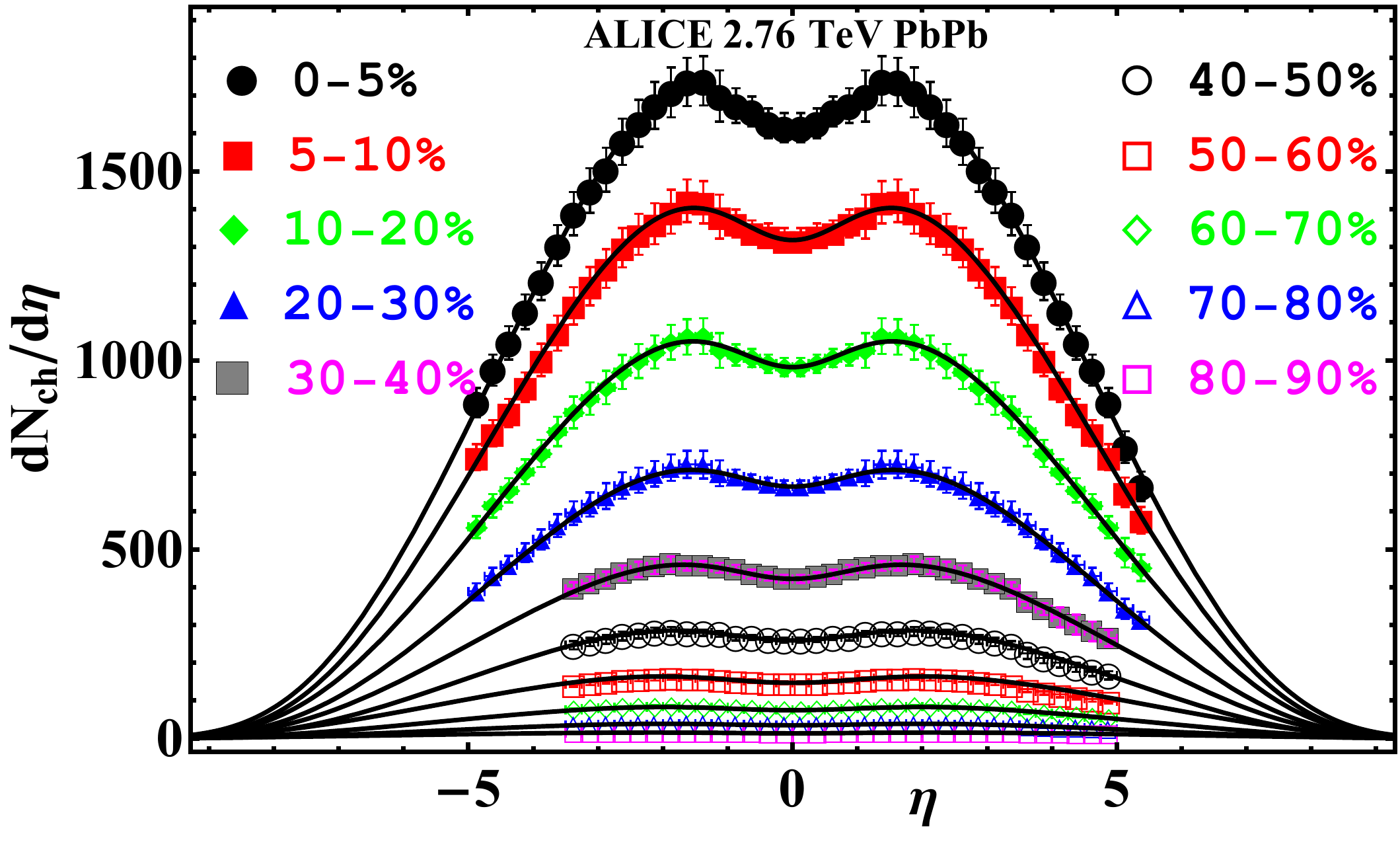}
\includegraphics[width=0.45\linewidth]{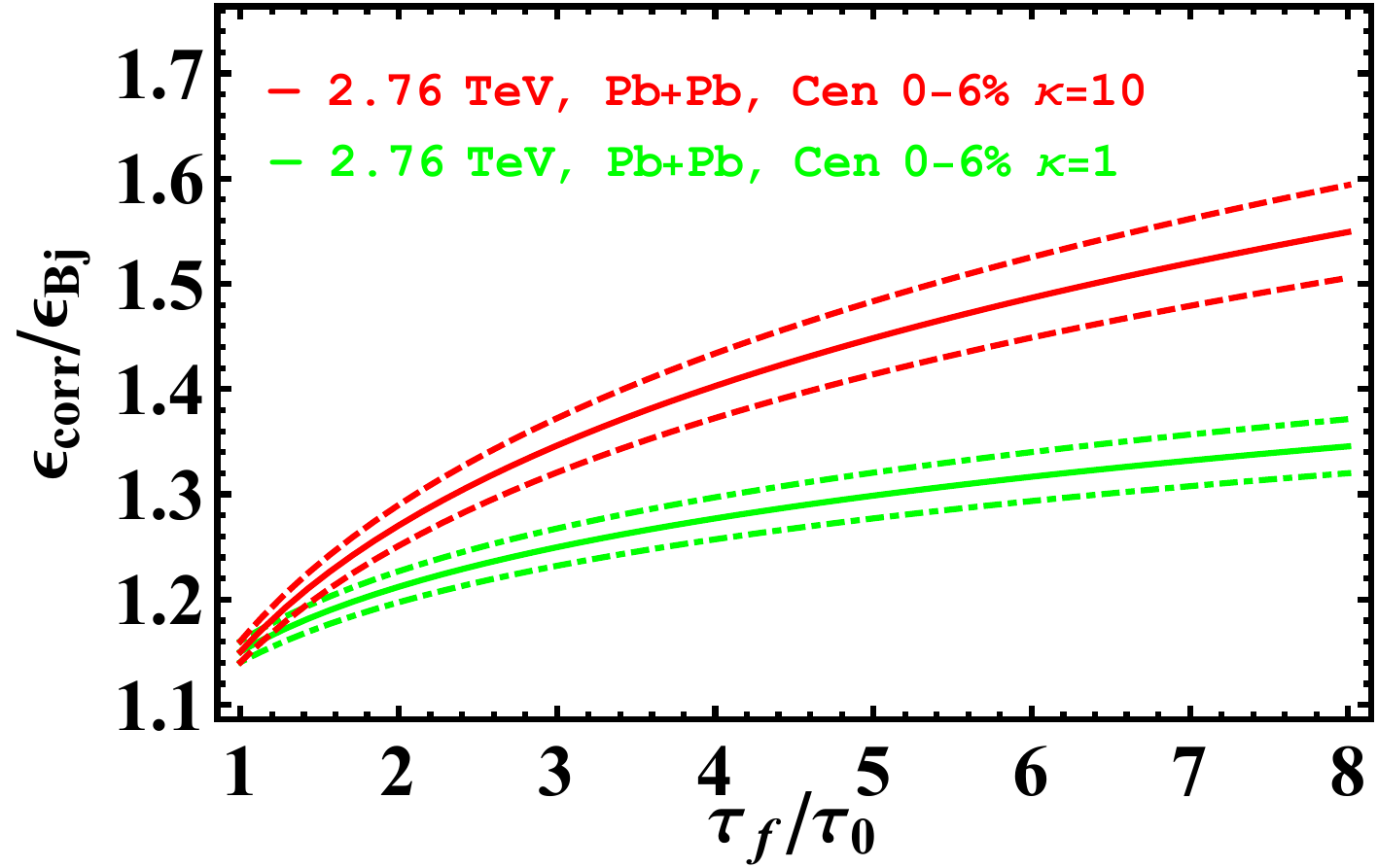}
\end{center}\vspace{-20pt}
\caption{(Color online) Plots in the left column show $dN_{\rm ch}/d\eta$ data measured by PHOBOS in
${130}$ GeV Au+Au collisions (first row), ${200}$ GeV Cu+Cu collisions (second row),
$200$ GeV Au+Au collisions (third row) and by ALICE in $2.76$ TeV Pb+Pb collisions (fourth row). These data
are compared to the hydro model result by the fit described in the paper.
Plots in the right column show the $\epsilon_{\rm corr}/\epsilon_{\rm Bj}$ correction factor,
as a function of the ratio of freeze-out time and thermalization time $\tau_{f}/\tau_{0}$,
for the most central collisions. Dashed lines represent the uncertainty.}\label{f:aafits}
\end{figure}

\begin{table}
\begin{center}
\begin{tabular}{l c c c c c }
\hline\hline
Centrality~[$\%$] & ~$\left.\frac{dN}{d\eta}\right\vert_{\eta=\eta_{0}}$ &~$\lambda$  &~$\sigma$  & $~T_{\rm eff}[{\rm GeV}]$ &~$\chi^{2}$/NDF    \\
\hline
~$0-6$    & 563.9$\pm$59.5   &~1.304$\pm$0.06   &~1.41$\pm$0.12  &~0.18$\pm$0.01   &55.0/51    \\
~$6-15$   & 437.6$\pm$41.2   &~1.255$\pm$0.08   &~1.37$\pm$0.16  &~0.18$\pm$0.01   &65.1/51    \\
~$15-25$  &  230.6$\pm$18.2  &~1.195$\pm$0.02   &~1.26$\pm$0.10  &~0.20$\pm$0.01   &77.8/51    \\
~$25-35$  &  152.5$\pm$13.1  &~1.178$\pm$0.03   &~1.28$\pm$0.04  &~0.21$\pm$0.01   &118.7/51   \\
~$35-45$  &  98.5$\pm$7.8    &~1.154$\pm$0.03   &~1.25$\pm$0.07  &~0.22$\pm$0.01   &121.6/51   \\
~$45-55$  &  67.8$\pm$5.5    &~1.132$\pm$0.04   &~1.16$\pm$0.10  &~0.23$\pm$0.01   &94.5/51    \\
\hline\hline
\end{tabular}
\end{center}
\caption{Fit parameters for 130 GeV Au+Au data, with their systematic uncertainties. Auxiliary values of $T_{f}=0.09$ GeV, $\bar{m}=0.24$ GeV have been utilized, based on Refs.~\cite{Csorgo:1995bi,Abelev:2008ab}}\label{t:130auaupars}
\end{table}

\begin{table}
\begin{center}
\begin{tabular}{l c c c c c }
\hline\hline
Centrality~[$\%$] & ~$\left.\frac{dN}{d\eta}\right\vert_{\eta=\eta_{0}}$ &~$\lambda$  &~$\sigma$  & $~T_{\rm eff}[{\rm GeV}]$ &~$\chi^{2}$/NDF    \\
\hline
~$0-6$    &642.6$\pm$61   &~1.239$\pm$0.045  &~1.41$\pm$0.10   &~0.18$\pm$0.01     &19.4/51    \\
~$6-15$   &498.5$\pm$45   &~1.214$\pm$0.045  &~1.40$\pm$0.08   &~0.19$\pm$0.01     &15.5/51    \\
~$15-25$  &347.5$\pm$32   &~1.189$\pm$0.045  &~1.36$\pm$0.12   &~0.20$\pm$0.01     &21.1/51    \\
~$25-35$  &243.2$\pm$22   &~1.159$\pm$0.026  &~1.30$\pm$0.08   &~0.21$\pm$0.01     &23.4/51    \\
~$35-45$  &151.5$\pm$15.5 &~1.137$\pm$0.023  &~1.24$\pm$0.03   &~0.22$\pm$0.01     &22.3/51    \\
~$45-55$  &91.8$\pm$8.8   &~1.133$\pm$0.026  &~1.28$\pm$0.10   &~0.22$\pm$0.01     &22.8/51    \\
\hline\hline
\end{tabular}
\end{center}
\caption{Fit parameters for 200 GeV Au+Au data, with their systematic uncertainties. Auxiliary values of $T_{f}=0.09$ GeV, $\bar{m}=0.24$ GeV have been utilized, based on Refs.~\cite{Csorgo:1995bi,Abelev:2008ab}}\label{t:200auaupars}
\end{table}

 \begin{table}
\begin{center}
\begin{tabular}{l c c c c c }
\hline\hline
Centrality~[$\%$] & ~$\left.\frac{dN}{d\eta}\right\vert_{\eta=\eta_{0}}$ &~$\lambda$  &~$\sigma$  & $~T_{\rm eff}[{\rm GeV}]$  &~$\chi^{2}$/NDF   \\
\hline
~$0-6$   &~179.5$\pm$17.5  &~1.176$\pm$0.025  &~1.28$\pm$0.07  &~0.20$\pm$0.01     &69.2/51    \\
~$6-15$  &~139.3$\pm$12.8  &~1.154$\pm$0.036  &~1.21$\pm$0.06  &~0.20$\pm$0.01     &42.2/51    \\
~$15-25$ &~99.4$\pm$9.6    &~1.133$\pm$0.030  &~1.16$\pm$0.06  &~0.21$\pm$0.01     &33.7/51    \\
~$25-35$ &~67.6$\pm$6.3    &~1.118$\pm$0.025  &~1.10$\pm$0.07  &~0.22$\pm$0.01     &34.5/51    \\
~$35-45$ &~45.3$\pm$3.95   &~1.113$\pm$0.029  &~1.12$\pm$0.06  &~0.22$\pm$0.01     &29.5/51    \\
~$45-55$ &~28.3$\pm$3.27   &~1.092$\pm$0.022  &~0.97$\pm$0.08  &~0.22$\pm$0.01     &33.0/51    \\
\hline
\end{tabular}
\end{center}
\caption{Fit parameters for 200 GeV Cu+Cu data, with their systematic uncertainties. Auxiliary values of $T_{f}=0.09$ GeV, $\bar{m}=0.24$ GeV have been utilized, based on Refs.~\cite{Csorgo:1995bi,Abelev:2008ab}}\label{t:200cucupars}
\end{table}

\begin{table}
\begin{center}
\begin{tabular}{l c c c c c }
\hline\hline
Centrality~[$\%$] & $\quad$$\left.\frac{dN}{d\eta}\right\vert_{\eta=\eta_{0}}$ &~$\lambda$  &~$\sigma$ & $~T_{\rm eff}[{\rm GeV}]$    &~$\chi^{2}$/NDF  \\ [0.5ex]
\hline
~$0-5$    & ~1615$\pm$39  &~1.075$\pm$0.005  &~0.82$\pm$0.05  &~0.29$\pm$0.01      &5.6/39     \\
~$5-10$   & ~1318$\pm$32  &~1.077$\pm$0.003  &~0.88$\pm$0.03  &~0.29$\pm$0.01      &4.2/39     \\
~$10-20$  & ~982$\pm$24   &~1.074$\pm$0.003  &~0.86$\pm$0.04  &~0.29$\pm$0.01      &3.9/39     \\
~$20-30$  & ~666$\pm$16   &~1.075$\pm$0.003  &~0.88$\pm$0.03  &~0.28$\pm$0.01      &3.0/39     \\
~$30-40$  &~422$\pm$11    &~1.073$\pm$0.001  &~0.89$\pm$0.04  &~0.27$\pm$0.02      &3.4/31     \\
~$40-50$  &~259.1$\pm$6.5 &~1.078$\pm$0.002  &~0.97$\pm$0.04  &~0.26$\pm$0.01      &4.2/31     \\
~$50-60$  &~147.1$\pm$3.6 &~1.080$\pm$0.001  &~1.02$\pm$0.03  &~0.25$\pm$0.01      &4.5/31     \\
~$60-70$  &~74.7$\pm$1.8  &~1.069$\pm$0.005  &~0.97$\pm$0.08  &~0.25$\pm$0.01      &9.8/31     \\
~$70-80$  &~34.8$\pm$0.86 &~1.069$\pm$0.003  &~1.01$\pm$0.07  &~0.26$\pm$0.01      &8.1/31     \\
~$80-90$  &~13.4$\pm$0.35 &~1.045$\pm$0.005  &~0.90$\pm$0.09  &~0.28$\pm$0.01      &10.2/31    \\
\hline\hline
\end{tabular}
\end{center}
\caption{Fit parameters for 2.76 TeV Pb+Pb data, with their systematic uncertainties. Auxiliary values of $T_{f}=0.17$ GeV, $\bar{m}=0.24$ GeV have been utilized, based on Refs.~\cite{Csorgo:1995bi,Abelev:2008ab}}\label{t:pbpbpars}
\end{table}

\begin{figure}
\begin{center}
\includegraphics[width=0.81\linewidth]{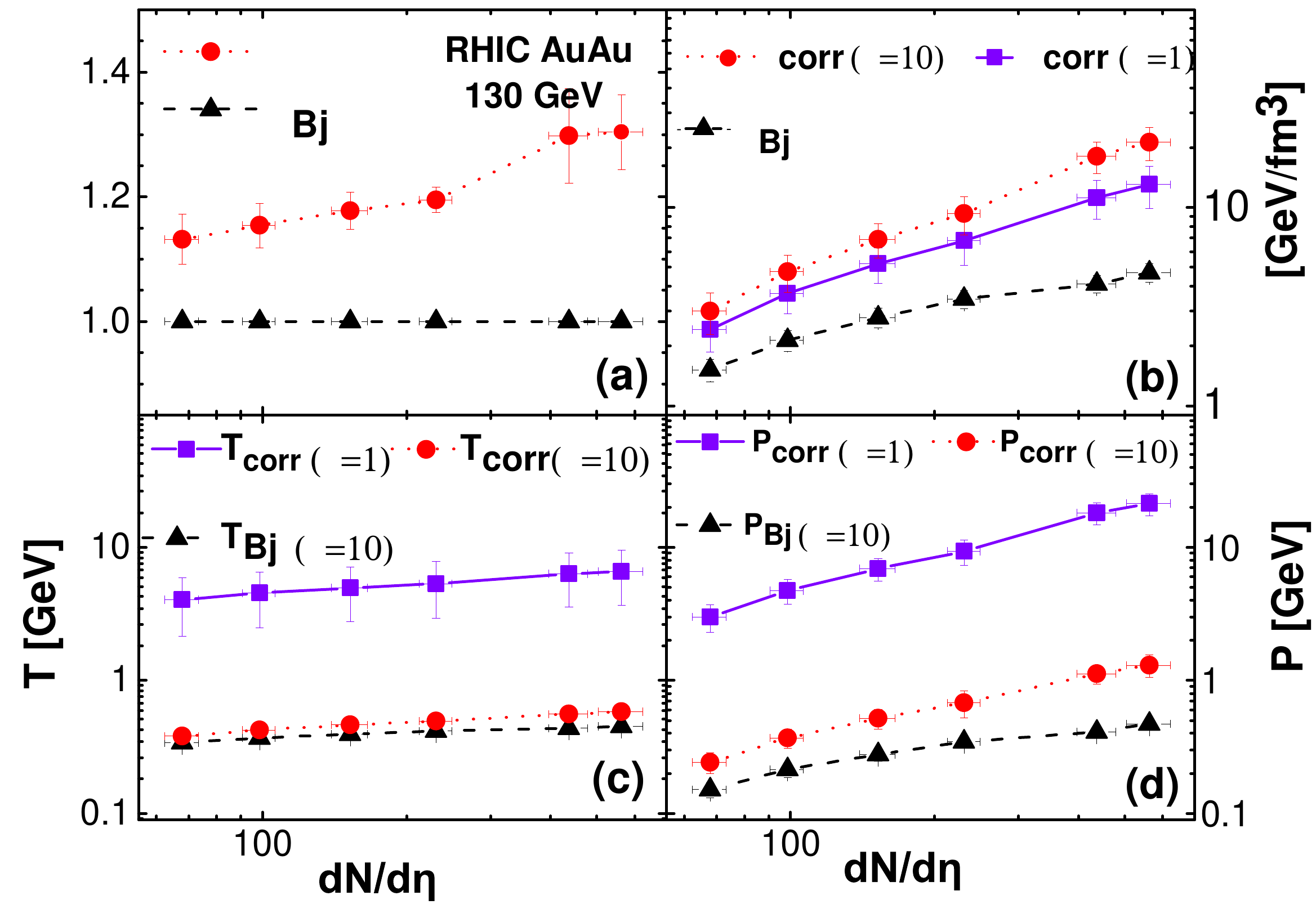}\vspace{-20pt}
\end{center}
\caption{(Color online) Acceleration parameter $\lambda$, initial energy density, temperature and pressure is indicated as a function of central multiplicity density and EoS parameter $\kappa$, for $\sqrt{s_{_{NN}}}=130$ GeV Au+Au collisions. Systematic uncertainties are also indicated, stemming from the determination of $\tau_{f}/\tau_{0}$, $\lambda$, $dN/d\eta$, as well as from the systematic uncertainties of the data.}\label{f:eps130auau}
\end{figure}

\begin{figure}
\begin{center}
\includegraphics[width=0.81\linewidth]{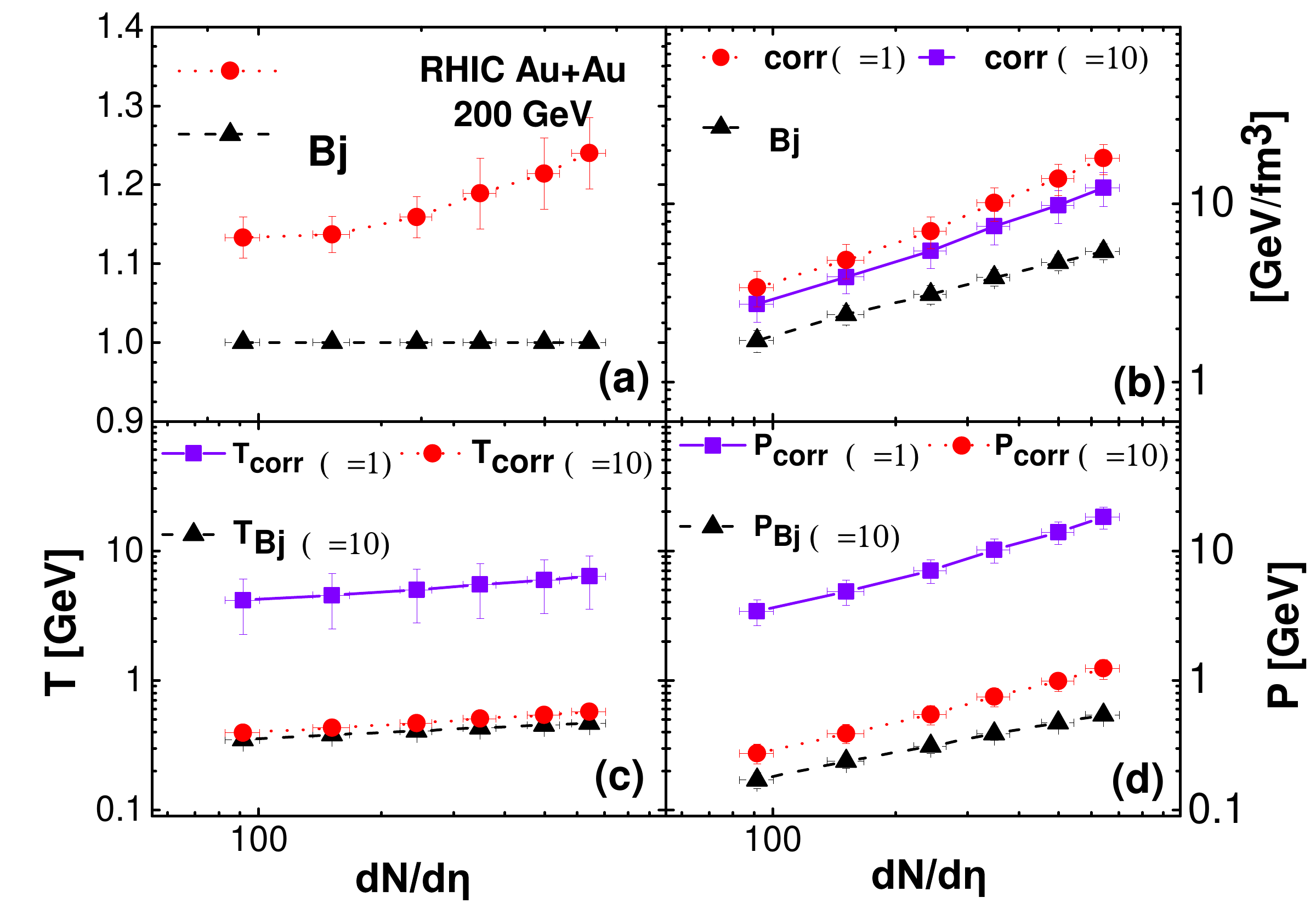}\vspace{-20pt}
\end{center}
\caption{(Color online) Acceleration parameter $\lambda$, initial energy density, temperature and pressure is indicated as a function of central multiplicity density and EoS parameter $\kappa$, for $\sqrt{s_{_{NN}}}=200$ GeV Au+Au collisions. Systematic uncertainties are also indicated, similarly to Fig.~\ref{f:eps130auau}.}\label{f:eps200auau}
\end{figure}

\begin{figure}
\begin{center}
\includegraphics[width=0.81\linewidth]{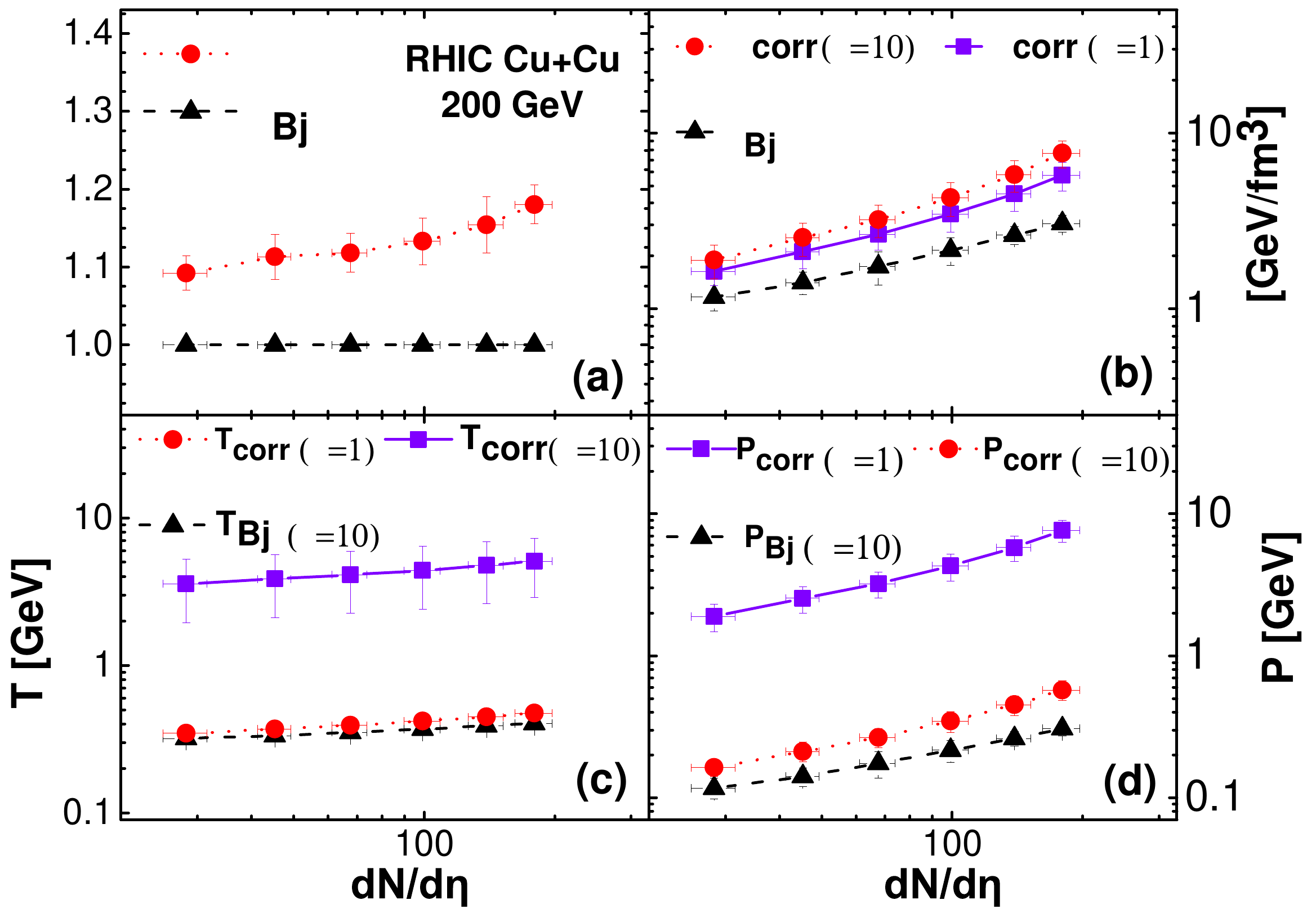}\vspace{-20pt}
\end{center}
\caption{(Color online) Acceleration parameter $\lambda$, initial energy density, temperature and pressure is indicated as a function of central multiplicity density and EoS parameter $\kappa$, for $\sqrt{s_{_{NN}}}=200$ GeV Cu+Cu collisions. Systematic uncertainties are also indicated, similarly to Fig.~\ref{f:eps130auau}.}\label{f:eps200cucu}
\end{figure}

\begin{figure}
\begin{center}
\includegraphics[width=0.81\linewidth]{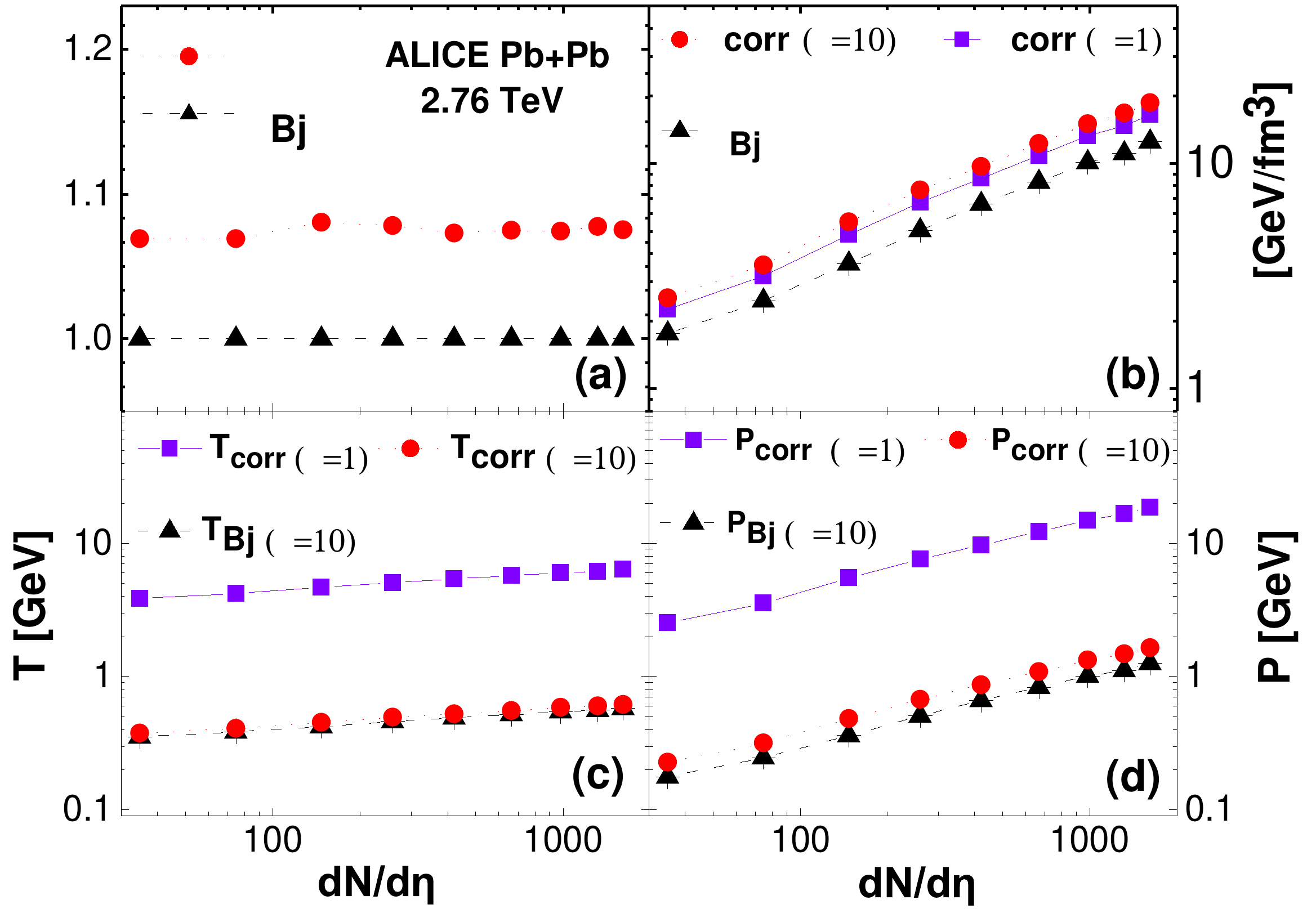}\vspace{-20pt}
\end{center}
\caption{(Color online) Acceleration parameter $\lambda$, initial energy density, temperature and pressure is indicated as a function of central multiplicity density and EoS parameter $\kappa$, for $\sqrt{s_{_{NN}}}=2.76$ TeV Pb+Pb collisions. Systematic uncertainties are also indicated, similarly to Fig.~\ref{f:eps130auau}.}\label{f:eps276pbpb}
\end{figure}

Let us then move to nucleus-nucleus collisions. We analyze RHIC PHOBOS $dN/d\eta$ data measured in $\sqrt{s_{_{NN}}}=130$ GeV Au+Au~\cite{Alver:2010ck}, 200 GeV Au+Au~\cite{Alver:2010ck} and 200 GeV Cu+Cu~\cite{Alver:2010ck} collisions of various centralities. We also analyze LHC ALICE $dN/d\eta$ data~\cite{Adam:2015kda} measured in $\sqrt{s_{_{NN}}}= 2.76$ TeV Pb+Pb collisions of various centralities. Fit results to these data are shown in Fig.~\ref{f:aafits}, and the fit parameters and properties are given in Tables~\ref{t:130auaupars}-\ref{t:pbpbpars}. Note that in this case and in all the subsequent cases, no statistical uncertainty was given experimentally, and also the point-by-point fluctuating part of the systematic uncertainty was not given. In order to be able to perform fits, we assumed an 3\% fluctuating systematic uncertainty, and used this value when minimizing the $\chi^2$ during the fits. We then used the full systematic uncertainties to estimate the systematic uncertainty of our parameters: we performed fits to datapoints shifted up and down by one unit of systematic uncertainty. In all the figures and tables, the parameter uncertainties represent this systematic uncertainty, as the statistical uncertainty was much smaller. Under these assumptions, all the analyzed data (all energies and centralities) are statistically well represented by the fitted curves, hence we may proceed to interpret the parameters. From the obtained acceleration values, we then calculate the energy density correction ratio $\epsilon_{\rm corr}/\epsilon_{\rm Bj}$, these are shown in the right column plots of Fig.~\ref{f:aafits}, as a function of $\tau_{f}/\tau_{0}$. In all cases, the initial energy density is strongly underestimated by the Bjorken model. The reason for this is the longitudinal acceleration, driven by pressure gradients and volume expansion~\cite{Csorgo:2006ax}. As shown in Figs.~\ref{f:eps130auau}-\ref{f:eps276pbpb} and Tables~\ref{t:130auaupars}-\ref{t:pppars}, the acceleration parameter $\lambda$ shows a clear trend: it decreases with collision energy from $1.304\pm0.002$ (130 GeV Au+Au, most central collisions) through $1.075\pm0.001$ (2.76 TeV Pb+Pb, most central collisions) to $1.067\pm0.001$ (8 TeV p+p). The multiplicity dependence of $\lambda$ is also similar in the RHIC cases: a roughly 10\% decrease in $\lambda$ is seen for mid-central collisions, as compared to the most central case. However, for the LHC Pb+Pb data, approximately constant values (around $\lambda\approx1.07$) are observed. Slightly lower $\lambda$ values are obtained from from 7 and 8 TeV p+p data -- acceleration seems to be much smaller at these high energies, in other words, almost perfect longitudinal Bjorken- or Hubble-flow is formed in these collisions. However, due to lack of centrality dependent $dN/d\eta$ data in p+p collisions at 7 and 8 TeV, for the analysis of these collisions we have assumed that $\lambda$ is approximately independent of mean multiplicity.

We can estimate the initial energy density of nucleus nucleus-nucleus collisions by assuming $\tau_{f}/\tau_{0}=6\pm2$ conservatively, based on Refs.~\cite{Csorgo:2007ea,Adare:2015bua}, using multiplicites from the $dN/d\eta$ data, and in case of ALICE data, using transverse energy distributions from Ref.~\cite{Adam:2016thv}. We may obtain the $S_{\perp}$ values from Refs.~\cite{Adare:2015bua,Abbas:2013taa} and utilize the EoS parameter $\kappa=10$, corresponding to $c_s\approx 0.32$~\cite{Lacey:2006bc,Borsanyi:2010cj,Csanad:2011jq}. We again estimate the temperature utilizing the $\epsilon \propto T^{4}$ relationship, similarly to Ref.~\cite{Csanad:2016add}, using $\epsilon_0=1$ GeV/fm$^3$ at $T_0=170$ MeV. We estimate the pressure using the EoS $\epsilon=\kappa p$. Alternatively, for comparison, we also utilize a very hard EoS of $\kappa=1$~\cite{Nagy:2007xn}. The results are shown in Figs.~\ref{f:eps130auau}-\ref{f:eps276pbpb}, with all the values given in Tables~\ref{t:130auaueps}-\ref{t:276pbpbeps}. We observe that the initial energy density is increasing with multiplicity (almost a factor of 3 from central to mid-peripheral) and collision energy. The Bjorken-estimate increases by roughly a factor of 3 when going from RHIC to LHC, but the corrected estimate, due to smaller acceleration, indicates a much smaller increase. While $\epsilon_{\rm corr}/\epsilon_{\rm Bj}$ may reach values of nearly 3 at RHIC, the change is only 20-30\% at the LHC. We also observe that the initial energy density is decreasing with system size, as seen from a 200 GeV Au+Au to Cu+Cu comparison. The multiplicity dependence of the initial temperature is qualitatively similar, albeit shows smaller changes, with values ranging from 0.35 GeV to 0.62 GeV. Pressure behaves similarly to the energy density, due to the linear EoS relationship.

\begin{table}
\begin{center}
\begin{tabular}{l c c c c  }
\hline\hline
Centrality~[$\%$] & $~\epsilon_{\rm Bj} [{\rm GeV/fm^{3}}]$ & $~\epsilon_{\rm corr} [{\rm GeV/fm^{3}}]$  & $~T_{\rm corr} [{\rm GeV}]$  &  $~P_{\rm corr} [{\rm GeV}]$ \\
\hline
~$0-6$            &  ~4.74$\pm$0.49  & 13.02$\pm$2.51    &~0.58$\pm$0.03     &~1.30$\pm$0.25             \\
~$6-15$          &  ~4.12$\pm$0.42  & 11.21$\pm$1.80    &~0.56$\pm$0.02     &~1.12$\pm$0.18          \\
~$15-25$              &  ~3.45$\pm$0.37    &  6.80$\pm$1.55     &~0.49$\pm$0.02     &~0.68$\pm$0.15           \\
~$25-35$        &  ~2.78$\pm$0.32    &  5.18$\pm$0.88     &~0.46$\pm$0.02     &~0.52$\pm$0.08            \\
~$35-45$        &  ~2.14$\pm$0.27  &  3.69$\pm$0.61     &~0.42$\pm$0.02     &~0.37$\pm$0.06          \\
~$45-55$        &  ~1.52$\pm$0.20  &  2.42$\pm$0.43   &~0.38$\pm$0.02     &~0.24$\pm$0.04             \\
\hline\hline
\end{tabular}
\end{center}
\caption{Thermodynamic quantities and their systematic uncertainties obtained by the hydrodynamic fits to $\sqrt{s_{_{NN}}}=130$ GeV Au+Au data from PHOBOS.}\label{t:130auaueps}
\end{table}

\begin{table}
\begin{center}
\begin{tabular}{l c c c c  }
\hline\hline
Centrality~[$\%$] & $~\epsilon_{\rm Bj} [{\rm GeV/fm^{3}}]$ & $~\epsilon_{\rm corr} [{\rm GeV/fm^{3}}]$  & $~T_{\rm corr} [{\rm GeV}]$  &  $~P_{\rm corr} [{\rm GeV}]$ \\
\hline
~$0-6$             &  ~5.42$\pm$0.54 &  12.33$\pm$2.12     &~0.57$\pm$0.03     &~1.23$\pm$0.21             \\
~$6-15$           &  ~4.70$\pm$0.48 &  9.85$\pm$1.60      &~0.54$\pm$0.02     &~0.99$\pm$0.16          \\
~$15-25$             &  ~3.87$\pm$0.41   &  7.49$\pm$1.23      &~0.51$\pm$0.02     &~0.75$\pm$0.12           \\
~$25-35$       &  ~3.10$\pm$0.37   &  5.43$\pm$0.91      &~0.47$\pm$0.02     &~0.54$\pm$0.09            \\
~$35-45$         &  ~2.40$\pm$0.30 &  3.89$\pm$0.65      &~0.43$\pm$0.02     &~0.39$\pm$0.06          \\
~$45-55$         &  ~1.71$\pm$0.24 &  2.74$\pm$0.48      &~0.39$\pm$0.02     &~0.27$\pm$0.05          \\
\hline\hline
\end{tabular}
\end{center}
\caption{Thermodynamic quantities and their systematic uncertainties obtained by the hydrodynamic fits to $\sqrt{s_{_{NN}}}=200$ GeV Au+Au data from PHOBOS.}\label{t:200auaueps}
\end{table}

\begin{table}
\begin{center}
\begin{tabular}{l c c c c  }
\hline\hline
Centrality~[$\%$] & $~\epsilon_{\rm Bj} [{\rm GeV/fm^{3}}]$ & $~\epsilon_{\rm corr} [{\rm GeV/fm^{3}}]$  & $~T_{\rm corr} [{\rm GeV}]$  &  $~P_{\rm corr} [{\rm GeV}]$ \\
\hline
~$0-6$              &  ~3.06$\pm$0.34   & 5.74$\pm$0.90    &~0.47$\pm$0.02     &~0.57$\pm$0.09             \\
~$6-15$            &  ~2.62$\pm$0.31   & 4.52$\pm$0.72    &~0.45$\pm$0.02     &~0.45$\pm$0.07          \\
~$15-25$          &  ~2.16$\pm$0.39   &  3.47$\pm$0.58    &~0.42$\pm$0.02     &~0.35$\pm$0.06           \\
~$25-35$          &  ~1.75$\pm$0.37   &  2.66$\pm$0.41    &~0.39$\pm$0.02     &~0.27$\pm$0.04            \\
~$35-45$          &  ~1.41$\pm$0.21   &  2.11$\pm$0.34    &~0.37$\pm$0.02     &~0.21$\pm$0.03          \\
~$45-55$          &  ~1.17$\pm$0.20   &  1.63$\pm$0.22    &~0.35$\pm$0.02     &~0.16$\pm$0.02             \\
\hline\hline
\end{tabular}
\end{center}
\caption{Thermodynamic quantities and their systematic uncertainties obtained by the hydrodynamic fits to $\sqrt{s_{_{NN}}}=200$ GeV Cu+Cu data from PHOBOS.}\label{t:200cucueps}
\end{table}

\begin{table}
\begin{center}
\begin{tabular}{l c c c c  }
\hline\hline
Centrality~[$\%$] & $~\epsilon_{\rm Bj} [{\rm GeV/fm^{3}}]$ & $~\epsilon_{\rm corr} [{\rm GeV/fm^{3}}]$  & $~T_{\rm corr} [{\rm GeV}]$  &  $~P_{\rm corr} [{\rm GeV}]$ \\
\hline
~~~$0-5$     &~12.53$\pm$0.44  &  16.50$\pm$0.81     &~0.62$\pm$0.02        &~~1.64$\pm$0.08         \\
~~~$5-10$    &~11.13$\pm$0.39 &  14.77$\pm$0.71     &~0.60$\pm$0.02        &~~1.48$\pm$0.07       \\
~~~$10-20$    &~10.14$\pm$0.34   &  13.30$\pm$0.64        &~0.58$\pm$0.02        &~~1.50$\pm$0.06        \\
~~~$20-30$   &~8.29$\pm$0.32  &  10.90$\pm$0.57     &~0.56$\pm$0.02        &~~1.27$\pm$0.06      \\
~~~$30-40$    &~6.61$\pm$0.25   &  8.62$\pm$0.46        &~0.52$\pm$0.02        &~~0.86$\pm$0.05      \\
~~~$40-50$     &~5.07$\pm$0.19  &  6.73$\pm$0.34      &~0.49$\pm$0.02        &~~0.67$\pm$0.03      \\
~~~$50-60$     &~3.61$\pm$0.15    &  4.84$\pm$0.26        &~0.45$\pm$0.02        &~~0.48$\pm$0.03        \\
~~~$60-70$   &~2.47$\pm$0.09  &  3.17$\pm$0.17      &~0.40$\pm$0.01        &~~0.32$\pm$0.02        \\
~~~$70-80$      &~1.76$\pm$0.10   & 2.26$\pm$0.16         &~0.38$\pm$0.01        &~~0.23$\pm$0.02       \\
\hline\hline
\end{tabular}
\end{center}
\caption{Thermodynamic quantities and their systematic uncertainties obtained by the hydrodynamic fits to $\sqrt{s_{_{NN}}}=2.76$ TeV Pb+Pb data from ALICE.}\label{t:276pbpbeps}
\end{table}

\section{Conclusions}\label{s:concl}
New results were shown on pseudorapidity distributions and the initial energy density estimate from the previous known exact accelerating solutions of hydrodynamics. The model result was successfully fitted to pseudorapidity densities from PHOBOS and ALICE. From these fits, we extracted a series of acceleration parameters $\lambda$ for different systems at RHIC and LHC energies. Taking the acceleration effect into account and refining the Bjorken model, we obtained an initial energy density estimation $\epsilon_{\rm corr}$ for different systems, significantly larger than the conventional Bjorken-estimate. For this estimate, we utilized transverse area values from MC Glauber simulations.  We found that there are clear trends in both collision energy and multiplicity: the acceleration is the largest in central collisions, and it decreases with increasing center of mass energy. The resulting corrected energy density estimate indicates that the energy density is increasing with collision energy and system size (nucleon size and centrality as well). We find that energy densities more than 10 GeV/fm$^3$ have been reached in central Au+Au collisions at RHIC and central Pb+Pb collisions at the LHC. We furthermore observe, that the calculated initial temperature and pressure depends strongly on the assumed equation of state, and hence these quantities shall be estimated based on penetrating probes (such as direct photons) or models that describe observables sensitive to the initial temperature. For now, we have utilized the average value for the speed of sound, $c_s\approx 0.32$, as determined from PHENIX measurements in $\sqrt{s_{_{NN}}}=200$ GeV Au+Au collisions~\cite{Lacey:2006bc,Borsanyi:2010cj,Csanad:2011jq}, that leads to a significant EoS dependent increase.

Our results indicate that the longitudinal expansion dynamics in heavy ion collisions at RHIC and LHC as well as proton-proton collisions at LHC energies can be described using the same exact, accelerating and finite solution of perfect fluid hydrodynamics. Our quantitative investigations also indicate that proton-proton collisions with about two times the average multiplicity can produce initial energy densities that are larger than 1 GeV/fm$^3$, the critical energy believed to be needed for the production of strongly interactive quark-gluon plasma. Hence one of the necessary conditions for quark-gluon plasma creation is satisfied in high multiplicity proton-proton collisions at LHC. The estimation of viscous corrections is currently under investigation but goes beyond the scope of the present paper.

\section*{Acknowledgements}
The authors thank Fu-Qiang Wang, Zi-Wei Lin and Xin-Nian Wang for useful discussions and suggestions about the initial energy density estimate and hydrodynamic evolution. Ze-Fang Jiang would like to thank David Zaslavsky for discussion of the computer program of this work and Xiangyu Wu for MC-Glauber simulation. This work was supported by the bilateral Chinese--Hungarian governmental project Grant No.T\'eT 12CN-1-2012-0016, by the Hungarian NKIFH grants No. FK-123842 and FK-123959, by the NNSF of China under grant No.11435004 and by the China CCNU PhD Fund 2016YBZZ100. M. Csan\'ad was supported by the J\'anos Bolyai Research Scholarship and the ÚNKP-17-4 New National Excellence Program of the Hungarian Ministry of Human Capacities. T. Cs\"org\H{o} was supported by the exchange programme of the Hungarian and the Ukrainian Academies of Sciences, grants NKM-82/2016 and NKM-92/2017 and by the EFOP 3.6.1-16-2016-00001 grant.

\bibliography{05AAFIT}

\begin{thebibliography}{39}
\expandafter\ifx\csname natexlab\endcsname\relax\def\natexlab#1{#1}\fi
\expandafter\ifx\csname bibnamefont\endcsname\relax
  \def\bibnamefont#1{#1}\fi
\expandafter\ifx\csname bibfnamefont\endcsname\relax
  \def\bibfnamefont#1{#1}\fi
\expandafter\ifx\csname citenamefont\endcsname\relax
  \def\citenamefont#1{#1}\fi
\expandafter\ifx\csname url\endcsname\relax
  \def\url#1{\texttt{#1}}\fi
\expandafter\ifx\csname urlprefix\endcsname\relax\def\urlprefix{URL }\fi
\providecommand{\bibinfo}[2]{#2}
\providecommand{\eprint}[2][]{\url{#2}}

\bibitem[{\citenamefont{Landau}(1953)}]{Landau:1953gs}
\bibinfo{author}{\bibfnamefont{L.~D.} \bibnamefont{Landau}},
  \bibinfo{journal}{Izv. Akad. Nauk Ser. Fiz.} \textbf{\bibinfo{volume}{17}},
  \bibinfo{pages}{51} (\bibinfo{year}{1953}).

\bibitem[{\citenamefont{Hwa}(1974)}]{Hwa:1974gn}
\bibinfo{author}{\bibfnamefont{R.~C.} \bibnamefont{Hwa}},
  \bibinfo{journal}{Phys. Rev.} \textbf{\bibinfo{volume}{D10}},
  \bibinfo{pages}{2260} (\bibinfo{year}{1974}).

\bibitem[{\citenamefont{Bjorken}(1983)}]{Bjorken:1982qr}
\bibinfo{author}{\bibfnamefont{J.~D.} \bibnamefont{Bjorken}},
  \bibinfo{journal}{Phys. Rev.} \textbf{\bibinfo{volume}{D27}},
  \bibinfo{pages}{140} (\bibinfo{year}{1983}).

\bibitem[{\citenamefont{Csan\'ad
  et~al.}(2004{\natexlab{a}})\citenamefont{Csan\'ad, Cs{\"o}rg\H{o}, and
  L{\"o}rstad}}]{Csanad:2003qa}
\bibinfo{author}{\bibfnamefont{M.}~\bibnamefont{Csan\'ad}},
  \bibinfo{author}{\bibfnamefont{T.}~\bibnamefont{Cs{\"o}rg\H{o}}},
  \bibnamefont{and}
  \bibinfo{author}{\bibfnamefont{B.}~\bibnamefont{L{\"o}rstad}},
  \bibinfo{journal}{Nucl. Phys.} \textbf{\bibinfo{volume}{A742}},
  \bibinfo{pages}{80} (\bibinfo{year}{2004}{\natexlab{a}}),
  \eprint{nucl-th/0310040}.

\bibitem[{\citenamefont{Csan\'ad
  et~al.}(2004{\natexlab{b}})\citenamefont{Csan\'ad, Cs{\"o}rg\H{o},
  L{\"o}rstad, and Ster}}]{Csanad:2004mm}
\bibinfo{author}{\bibfnamefont{M.}~\bibnamefont{Csan\'ad}},
  \bibinfo{author}{\bibfnamefont{T.}~\bibnamefont{Cs{\"o}rg\H{o}}},
  \bibinfo{author}{\bibfnamefont{B.}~\bibnamefont{L{\"o}rstad}},
  \bibnamefont{and} \bibinfo{author}{\bibfnamefont{A.}~\bibnamefont{Ster}},
  \bibinfo{journal}{J. Phys.} \textbf{\bibinfo{volume}{G30}},
  \bibinfo{pages}{S1079} (\bibinfo{year}{2004}{\natexlab{b}}),
  \eprint{nucl-th/0403074}.

\bibitem[{\citenamefont{Agababyan et~al.}(1998)}]{Agababyan:1997wd}
\bibinfo{author}{\bibfnamefont{N.~M.} \bibnamefont{Agababyan}}
  \bibnamefont{et~al.} (\bibinfo{collaboration}{EHS/NA22}),
  \bibinfo{journal}{Phys. Lett.} \textbf{\bibinfo{volume}{B422}},
  \bibinfo{pages}{359} (\bibinfo{year}{1998}), \eprint{hep-ex/9711009}.

\bibitem[{\citenamefont{Cs\"org\H{o} et~al.}(2005)\citenamefont{Cs\"org\H{o},
  Csan\'ad, L\"orstad, and Ster}}]{Csorgo:2004id}
\bibinfo{author}{\bibfnamefont{T.}~\bibnamefont{Cs\"org\H{o}}},
  \bibinfo{author}{\bibfnamefont{M.}~\bibnamefont{Csan\'ad}},
  \bibinfo{author}{\bibfnamefont{B.}~\bibnamefont{L\"orstad}},
  \bibnamefont{and} \bibinfo{author}{\bibfnamefont{A.}~\bibnamefont{Ster}},
  \bibinfo{journal}{Acta Phys. Hung.} \textbf{\bibinfo{volume}{A24}},
  \bibinfo{pages}{139} (\bibinfo{year}{2005}), \eprint{hep-ph/0406042}.

\bibitem[{\citenamefont{Heinz and Kolb}(2002)}]{Heinz:2001xi}
\bibinfo{author}{\bibfnamefont{U.~W.} \bibnamefont{Heinz}} \bibnamefont{and}
  \bibinfo{author}{\bibfnamefont{P.~F.} \bibnamefont{Kolb}},
  \bibinfo{journal}{Nucl. Phys.} \textbf{\bibinfo{volume}{A702}},
  \bibinfo{pages}{269} (\bibinfo{year}{2002}), \eprint{hep-ph/0111075}.

\bibitem[{\citenamefont{Bass et~al.}(2009)\citenamefont{Bass, Gale, Majumder,
  Nonaka, Qin, Renk, and Ruppert}}]{Bass:2008rv}
\bibinfo{author}{\bibfnamefont{S.~A.} \bibnamefont{Bass}},
  \bibinfo{author}{\bibfnamefont{C.}~\bibnamefont{Gale}},
  \bibinfo{author}{\bibfnamefont{A.}~\bibnamefont{Majumder}},
  \bibinfo{author}{\bibfnamefont{C.}~\bibnamefont{Nonaka}},
  \bibinfo{author}{\bibfnamefont{G.-Y.} \bibnamefont{Qin}},
  \bibinfo{author}{\bibfnamefont{T.}~\bibnamefont{Renk}}, \bibnamefont{and}
  \bibinfo{author}{\bibfnamefont{J.}~\bibnamefont{Ruppert}},
  \bibinfo{journal}{Phys. Rev.} \textbf{\bibinfo{volume}{C79}},
  \bibinfo{pages}{024901} (\bibinfo{year}{2009}), \eprint{0808.0908}.

\bibitem[{\citenamefont{Huovinen et~al.}(2001)\citenamefont{Huovinen, Kolb,
  Heinz, Ruuskanen, and Voloshin}}]{Huovinen:2001cy}
\bibinfo{author}{\bibfnamefont{P.}~\bibnamefont{Huovinen}},
  \bibinfo{author}{\bibfnamefont{P.~F.} \bibnamefont{Kolb}},
  \bibinfo{author}{\bibfnamefont{U.~W.} \bibnamefont{Heinz}},
  \bibinfo{author}{\bibfnamefont{P.~V.} \bibnamefont{Ruuskanen}},
  \bibnamefont{and} \bibinfo{author}{\bibfnamefont{S.~A.}
  \bibnamefont{Voloshin}}, \bibinfo{journal}{Phys. Lett.}
  \textbf{\bibinfo{volume}{B503}}, \bibinfo{pages}{58} (\bibinfo{year}{2001}),
  \eprint{hep-ph/0101136}.

\bibitem[{\citenamefont{Kolb and Heinz}(2003)}]{Kolb:2003dz}
\bibinfo{author}{\bibfnamefont{P.~F.} \bibnamefont{Kolb}} \bibnamefont{and}
  \bibinfo{author}{\bibfnamefont{U.~W.} \bibnamefont{Heinz}}
  (\bibinfo{year}{2003}), \eprint{nucl-th/0305084}.

\bibitem[{\citenamefont{Song et~al.}(2011)\citenamefont{Song, Bass, Heinz,
  Hirano, and Shen}}]{Song:2010mg}
\bibinfo{author}{\bibfnamefont{H.}~\bibnamefont{Song}},
  \bibinfo{author}{\bibfnamefont{S.~A.} \bibnamefont{Bass}},
  \bibinfo{author}{\bibfnamefont{U.}~\bibnamefont{Heinz}},
  \bibinfo{author}{\bibfnamefont{T.}~\bibnamefont{Hirano}}, \bibnamefont{and}
  \bibinfo{author}{\bibfnamefont{C.}~\bibnamefont{Shen}},
  \bibinfo{journal}{Phys. Rev. Lett.} \textbf{\bibinfo{volume}{106}},
  \bibinfo{pages}{192301} (\bibinfo{year}{2011}), \eprint{1011.2783}.

\bibitem[{\citenamefont{Pang et~al.}(2016)\citenamefont{Pang, Petersen, Wang,
  and Wang}}]{Pang:2016igs}
\bibinfo{author}{\bibfnamefont{L.-G.} \bibnamefont{Pang}},
  \bibinfo{author}{\bibfnamefont{H.}~\bibnamefont{Petersen}},
  \bibinfo{author}{\bibfnamefont{Q.}~\bibnamefont{Wang}}, \bibnamefont{and}
  \bibinfo{author}{\bibfnamefont{X.-N.} \bibnamefont{Wang}},
  \bibinfo{journal}{Phys. Rev. Lett.} \textbf{\bibinfo{volume}{117}},
  \bibinfo{pages}{192301} (\bibinfo{year}{2016}), \eprint{1605.04024}.

\bibitem[{\citenamefont{Jiang et~al.}(2015)\citenamefont{Jiang, Zhang, Zhang,
  and Deng}}]{Jiang:2015apa}
\bibinfo{author}{\bibfnamefont{Z.~J.} \bibnamefont{Jiang}},
  \bibinfo{author}{\bibfnamefont{Y.}~\bibnamefont{Zhang}},
  \bibinfo{author}{\bibfnamefont{H.~L.} \bibnamefont{Zhang}}, \bibnamefont{and}
  \bibinfo{author}{\bibfnamefont{H.~P.} \bibnamefont{Deng}},
  \bibinfo{journal}{Nucl. Phys.} \textbf{\bibinfo{volume}{A941}},
  \bibinfo{pages}{188} (\bibinfo{year}{2015}), \eprint{1510.00160}.

\bibitem[{\citenamefont{Chen et~al.}(2016)\citenamefont{Chen, Pang, Stoecker,
  Luo, Wang, and Wang}}]{Chen:2016owv}
\bibinfo{author}{\bibfnamefont{W.}~\bibnamefont{Chen}},
  \bibinfo{author}{\bibfnamefont{L.-G.} \bibnamefont{Pang}},
  \bibinfo{author}{\bibfnamefont{H.}~\bibnamefont{Stoecker}},
  \bibinfo{author}{\bibfnamefont{T.}~\bibnamefont{Luo}},
  \bibinfo{author}{\bibfnamefont{E.}~\bibnamefont{Wang}}, \bibnamefont{and}
  \bibinfo{author}{\bibfnamefont{X.-N.} \bibnamefont{Wang}},
  \bibinfo{journal}{Nucl. Phys.} \textbf{\bibinfo{volume}{A956}},
  \bibinfo{pages}{605} (\bibinfo{year}{2016}).

\bibitem[{\citenamefont{Abelev et~al.}(2009)}]{Abelev:2008ab}
\bibinfo{author}{\bibfnamefont{B.~I.} \bibnamefont{Abelev}}
  \bibnamefont{et~al.} (\bibinfo{collaboration}{STAR}), \bibinfo{journal}{Phys.
  Rev.} \textbf{\bibinfo{volume}{C79}}, \bibinfo{pages}{034909}
  (\bibinfo{year}{2009}), \eprint{0808.2041}.

\bibitem[{\citenamefont{Adare et~al.}(2016)}]{Adare:2015bua}
\bibinfo{author}{\bibfnamefont{A.}~\bibnamefont{Adare}} \bibnamefont{et~al.}
  (\bibinfo{collaboration}{PHENIX}), \bibinfo{journal}{Phys. Rev.}
  \textbf{\bibinfo{volume}{C93}}, \bibinfo{pages}{024901}
  (\bibinfo{year}{2016}), \eprint{1509.06727}.

\bibitem[{\citenamefont{Adam et~al.}(2016{\natexlab{a}})}]{Adam:2015kda}
\bibinfo{author}{\bibfnamefont{J.}~\bibnamefont{Adam}} \bibnamefont{et~al.}
  (\bibinfo{collaboration}{ALICE}), \bibinfo{journal}{Phys. Lett.}
  \textbf{\bibinfo{volume}{B754}}, \bibinfo{pages}{373}
  (\bibinfo{year}{2016}{\natexlab{a}}), \eprint{1509.07299}.

\bibitem[{\citenamefont{Cs{\"o}rg{\H{o}} and
  L{\"o}rstad}(1996)}]{Csorgo:1995bi}
\bibinfo{author}{\bibfnamefont{T.}~\bibnamefont{Cs{\"o}rg{\H{o}}}}
  \bibnamefont{and}
  \bibinfo{author}{\bibfnamefont{B.}~\bibnamefont{L{\"o}rstad}},
  \bibinfo{journal}{Phys. Rev.} \textbf{\bibinfo{volume}{C54}},
  \bibinfo{pages}{1390} (\bibinfo{year}{1996}), \eprint{hep-ph/9509213}.

\bibitem[{\citenamefont{Cs{\"o}rg{\H{o}}
  et~al.}(2008)\citenamefont{Cs{\"o}rg{\H{o}}, Nagy, and
  Csan{\'a}d}}]{Csorgo:2006ax}
\bibinfo{author}{\bibfnamefont{T.}~\bibnamefont{Cs{\"o}rg{\H{o}}}},
  \bibinfo{author}{\bibfnamefont{M.~I.} \bibnamefont{Nagy}}, \bibnamefont{and}
  \bibinfo{author}{\bibfnamefont{M.}~\bibnamefont{Csan{\'a}d}},
  \bibinfo{journal}{Phys. Lett.} \textbf{\bibinfo{volume}{B663}},
  \bibinfo{pages}{306} (\bibinfo{year}{2008}), \eprint{nucl-th/0605070}.

\bibitem[{\citenamefont{Nagy et~al.}(2008)\citenamefont{Nagy, Cs{\"o}rg{\H{o}},
  and Csan{\'a}d}}]{Nagy:2007xn}
\bibinfo{author}{\bibfnamefont{M.~I.} \bibnamefont{Nagy}},
  \bibinfo{author}{\bibfnamefont{T.}~\bibnamefont{Cs{\"o}rg{\H{o}}}},
  \bibnamefont{and}
  \bibinfo{author}{\bibfnamefont{M.}~\bibnamefont{Csan{\'a}d}},
  \bibinfo{journal}{Phys. Rev.} \textbf{\bibinfo{volume}{C77}},
  \bibinfo{pages}{024908} (\bibinfo{year}{2008}), \eprint{0709.3677}.

\bibitem[{\citenamefont{Csan{\'a}d and
  Cs{\"o}rg{\H{o}}}(2017)}]{Csanad:2013lba}
\bibinfo{author}{\bibfnamefont{M.}~\bibnamefont{Csan{\'a}d}} \bibnamefont{and}
  \bibinfo{author}{\bibfnamefont{T.}~\bibnamefont{Cs{\"o}rg{\H{o}}}},
  \bibinfo{journal}{JCEGI} \textbf{\bibinfo{volume}{5}}, \bibinfo{pages}{97}
  (\bibinfo{year}{2017}), \eprint{1307.2082}.

\bibitem[{\citenamefont{Csan{\'a}d et~al.}(2017)\citenamefont{Csan{\'a}d,
  Cs{\"o}rg{\H{o}}, Jiang, and Yang}}]{Csanad:2016add}
\bibinfo{author}{\bibfnamefont{M.}~\bibnamefont{Csan{\'a}d}},
  \bibinfo{author}{\bibfnamefont{T.}~\bibnamefont{Cs{\"o}rg{\H{o}}}},
  \bibinfo{author}{\bibfnamefont{Z.-F.} \bibnamefont{Jiang}}, \bibnamefont{and}
  \bibinfo{author}{\bibfnamefont{C.-B.} \bibnamefont{Yang}},
  \bibinfo{journal}{Universe} \textbf{\bibinfo{volume}{3}}, \bibinfo{pages}{9}
  (\bibinfo{year}{2017}), \eprint{1609.07176}.

\bibitem[{\citenamefont{Cooper and Frye}(1974)}]{Cooper:1974mv}
\bibinfo{author}{\bibfnamefont{F.}~\bibnamefont{Cooper}} \bibnamefont{and}
  \bibinfo{author}{\bibfnamefont{G.}~\bibnamefont{Frye}},
  \bibinfo{journal}{Phys. Rev.} \textbf{\bibinfo{volume}{D10}},
  \bibinfo{pages}{186} (\bibinfo{year}{1974}).

\bibitem[{\citenamefont{Abbas et~al.}(2013)}]{Abbas:2013taa}
\bibinfo{author}{\bibfnamefont{E.}~\bibnamefont{Abbas}} \bibnamefont{et~al.}
  (\bibinfo{collaboration}{ALICE}), \bibinfo{journal}{JINST}
  \textbf{\bibinfo{volume}{8}}, \bibinfo{pages}{P10016} (\bibinfo{year}{2013}),
  \eprint{1306.3130}.

\bibitem[{\citenamefont{Adam et~al.}(2016{\natexlab{b}})}]{Adam:2016thv}
\bibinfo{author}{\bibfnamefont{J.}~\bibnamefont{Adam}} \bibnamefont{et~al.}
  (\bibinfo{collaboration}{ALICE}), \bibinfo{journal}{Phys. Rev.}
  \textbf{\bibinfo{volume}{C94}}, \bibinfo{pages}{034903}
  (\bibinfo{year}{2016}{\natexlab{b}}), \eprint{1603.04775}.

\bibitem[{\citenamefont{Alver et~al.}(2008)}]{Alver:2008zza}
\bibinfo{author}{\bibfnamefont{B.}~\bibnamefont{Alver}} \bibnamefont{et~al.},
  \bibinfo{journal}{Phys. Rev.} \textbf{\bibinfo{volume}{C77}},
  \bibinfo{pages}{014906} (\bibinfo{year}{2008}), \eprint{0711.3724}.

\bibitem[{\citenamefont{Gyulassy and Matsui}(1984)}]{Gyulassy:1983ub}
\bibinfo{author}{\bibfnamefont{M.}~\bibnamefont{Gyulassy}} \bibnamefont{and}
  \bibinfo{author}{\bibfnamefont{T.}~\bibnamefont{Matsui}},
  \bibinfo{journal}{Phys. Rev.} \textbf{\bibinfo{volume}{D29}},
  \bibinfo{pages}{419} (\bibinfo{year}{1984}).

\bibitem[{\citenamefont{Alver et~al.}(2011)}]{Alver:2010ck}
\bibinfo{author}{\bibfnamefont{B.}~\bibnamefont{Alver}} \bibnamefont{et~al.}
  (\bibinfo{collaboration}{PHOBOS}), \bibinfo{journal}{Phys. Rev.}
  \textbf{\bibinfo{volume}{C83}}, \bibinfo{pages}{024913}
  (\bibinfo{year}{2011}), \eprint{1011.1940}.

\bibitem[{\citenamefont{Khachatryan et~al.}(2010)}]{Khachatryan:2010us}
\bibinfo{author}{\bibfnamefont{V.}~\bibnamefont{Khachatryan}}
  \bibnamefont{et~al.} (\bibinfo{collaboration}{CMS}), \bibinfo{journal}{Phys.
  Rev. Lett.} \textbf{\bibinfo{volume}{105}}, \bibinfo{pages}{022002}
  (\bibinfo{year}{2010}), \eprint{1005.3299}.

\bibitem[{\citenamefont{Antchev et~al.}(2012)}]{Aspell:2012ux}
\bibinfo{author}{\bibfnamefont{G.}~\bibnamefont{Antchev}} \bibnamefont{et~al.}
  (\bibinfo{collaboration}{TOTEM}), \bibinfo{journal}{EPL}
  \textbf{\bibinfo{volume}{98}}, \bibinfo{pages}{31002} (\bibinfo{year}{2012}),
  \eprint{1205.4105}.

\bibitem[{\citenamefont{Chatrchyan et~al.}(2014)}]{Chatrchyan:2014qka}
\bibinfo{author}{\bibfnamefont{S.}~\bibnamefont{Chatrchyan}}
  \bibnamefont{et~al.} (\bibinfo{collaboration}{CMS, TOTEM}),
  \bibinfo{journal}{Eur. Phys. J.} \textbf{\bibinfo{volume}{C74}},
  \bibinfo{pages}{3053} (\bibinfo{year}{2014}), \eprint{1405.0722}.

\bibitem[{\citenamefont{Antchev et~al.}(2015)}]{Antchev:2014lez}
\bibinfo{author}{\bibfnamefont{G.}~\bibnamefont{Antchev}} \bibnamefont{et~al.}
  (\bibinfo{collaboration}{TOTEM}), \bibinfo{journal}{Eur. Phys. J.}
  \textbf{\bibinfo{volume}{C75}}, \bibinfo{pages}{126} (\bibinfo{year}{2015}),
  \eprint{1411.4963}.

\bibitem[{\citenamefont{Huovinen and Petreczky}(2010)}]{Huovinen:2009yb}
\bibinfo{author}{\bibfnamefont{P.}~\bibnamefont{Huovinen}} \bibnamefont{and}
  \bibinfo{author}{\bibfnamefont{P.}~\bibnamefont{Petreczky}},
  \bibinfo{journal}{Nucl. Phys.} \textbf{\bibinfo{volume}{A837}},
  \bibinfo{pages}{26} (\bibinfo{year}{2010}), \eprint{0912.2541}.

\bibitem[{\citenamefont{Bazavov et~al.}(2009)}]{Bazavov:2009zn}
\bibinfo{author}{\bibfnamefont{A.}~\bibnamefont{Bazavov}} \bibnamefont{et~al.},
  \bibinfo{journal}{Phys. Rev.} \textbf{\bibinfo{volume}{D80}},
  \bibinfo{pages}{014504} (\bibinfo{year}{2009}), \eprint{0903.4379}.

\bibitem[{\citenamefont{Lacey et~al.}(2007)}]{Lacey:2006bc}
\bibinfo{author}{\bibfnamefont{R.~A.} \bibnamefont{Lacey}}
  \bibnamefont{et~al.}, \bibinfo{journal}{Phys. Rev. Lett.}
  \textbf{\bibinfo{volume}{98}}, \bibinfo{pages}{092301}
  (\bibinfo{year}{2007}), \eprint{nucl-ex/0609025}.

\bibitem[{\citenamefont{Bors\'anyi et~al.}(2010)}]{Borsanyi:2010cj}
\bibinfo{author}{\bibfnamefont{S.}~\bibnamefont{Bors\'anyi}}
  \bibnamefont{et~al.}, \bibinfo{journal}{JHEP} \textbf{\bibinfo{volume}{11}},
  \bibinfo{pages}{077} (\bibinfo{year}{2010}), \eprint{1007.2580}.

\bibitem[{\citenamefont{Csan\'ad and M\'ajer}(2012)}]{Csanad:2011jq}
\bibinfo{author}{\bibfnamefont{M.}~\bibnamefont{Csan\'ad}} \bibnamefont{and}
  \bibinfo{author}{\bibfnamefont{I.}~\bibnamefont{M\'ajer}},
  \bibinfo{journal}{Central Eur.J.Phys.} \textbf{\bibinfo{volume}{10}},
  \bibinfo{pages}{850} (\bibinfo{year}{2012}), \eprint{1101.1279}.

\bibitem[{\citenamefont{Cs{\"o}rg\H{o}
  et~al.}(2007)\citenamefont{Cs{\"o}rg\H{o}, Nagy, and
  Csan\'ad}}]{Csorgo:2007ea}
\bibinfo{author}{\bibfnamefont{T.}~\bibnamefont{Cs{\"o}rg\H{o}}},
  \bibinfo{author}{\bibfnamefont{M.~I.} \bibnamefont{Nagy}}, \bibnamefont{and}
  \bibinfo{author}{\bibfnamefont{M.}~\bibnamefont{Csan\'ad}},
  \bibinfo{journal}{Braz. J. Phys.} \textbf{\bibinfo{volume}{37}},
  \bibinfo{pages}{723} (\bibinfo{year}{2007}), \eprint{nucl-th/0702043}.

\end{thebibliography}

\end{document}